\documentclass[fleqn,usenatbib]{rasti}
\usepackage{newtxtext,newtxmath}
\usepackage[british]{babel}

\usepackage[T1]{fontenc}

\usepackage{booktabs}
\usepackage{subcaption}
\usepackage{placeins}
\DeclareRobustCommand{\VAN}[3]{#2}
\let\VANthebibliography\thebibliography
\def\thebibliography{\DeclareRobustCommand{\VAN}[3]{##3}\VANthebibliography}

\usepackage{float}
\usepackage{graphicx}	
\usepackage{xcolor}
\usepackage{amsmath}	

\usepackage{amssymb}	
\usepackage[acronym]{glossaries}
\glsdisablehyper
\usepackage{siunitx}
\usepackage{xurl}
\usepackage{hyperref}
\usepackage[capitalise,nameinlink]{cleveref}

\addto\extrasbritish{%
  }

\newacronym{fcn}{FCN}{Fully Connected Neural Network}
\newacronym{cnn}{CNN}{Convolutional Neural Network}
\newacronym{wgan}{wGAN}{Wasserstein Generative Adversarial Network}
\newacronym{vit}{ViT}{Vision Transformer}

\title[wGAN-supported augmentation]{Morphological Classification of Radio Galaxies with wGAN-supported Augmentation}

\author[Lennart Rustige et al.]{Lennart Rustige$^{1,2}$\thanks{E-mail: lennart.rustige@desy.de},
Janis Kummer$^{1,3}$\textsuperscript{\thanks{E-mail: janis.kummer@uni-hamburg.de}},
Florian Griese$^{1,4,5}$\textsuperscript{\thanks{E-mail: florian.griese@tuhh.de}},
Kerstin Borras$^{2,6}$,
Marcus~Br\"uggen$^{3}$,\newauthor
Patrick L. S. Connor$^{1,7}$,
Frank Gaede$^{2}$,
Gregor Kasieczka$^{7}$,
Tobias Knopp$^{4,5}$
and Peter Schleper$^{7}$
\\
$^{1}$ Center for Data and Computing in Natural Sciences (CDCS), Notkestrasse 9, D-22607 Hamburg, Germany\\
$^{2}$ Deutsches Elektronen-Synchrotron DESY, Notkestrasse 85, D-22607 Hamburg, Germany\\
$^{3}$ Universit\"at Hamburg, Hamburger Sternwarte, Gojenbergsweg 112, D-21029 Hamburg, Germany\\
$^{4}$ Section for Biomedical Imaging, University Medical Center Hamburg-Eppendorf, D-20246 Hamburg, Germany\\
$^{5}$ Institute for Biomedical Imaging, Hamburg University of Technology, D-21073 Hamburg, Germany\\
$^{6}$ RWTH Aachen University, Templergraben 55, D-52062 Aachen, Germany\\
$^{7}$ Universit\"at Hamburg, Institut für Experimentalphysik, Luruper Chaussee 149, D-22761
Hamburg, Germany\\
}

\date{Accepted 19/05/2023. Received 25/04/2023; in original form 16/12/2022}

\pubyear{2023}

\begin{document}
\label{firstpage}
\pagerange{\pageref{firstpage}--\pageref{lastpage}}
\maketitle

\begin{abstract}
Machine learning techniques that perform morphological classification of astronomical sources often suffer from a scarcity of labelled training data.
Here, we focus on the case of supervised deep learning models for the morphological classification of radio galaxies, which is particularly topical for the forthcoming large radio surveys. We demonstrate the use of generative models, specifically \glspl{wgan}, to generate data for different classes of radio galaxies. 
Further, we study the impact of augmenting the training data with images from our \gls{wgan} on three different classification architectures. We find that  this technique makes it possible to improve models for the morphological classification of radio galaxies. A simple \gls{fcn} benefits most from including generated images into the training set, with a considerable improvement of its classification accuracy. In addition, we find it is more difficult to improve complex classifiers. The classification performance of a \gls{cnn} can be improved slightly. However, this is not the  case for a \gls{vit}.
\end{abstract}

\glsresetall[main,acronym]




\section{Introduction}

Radio galaxies are galaxies that emit a large fraction of their electromagnetic output in the radio band. The structures visible in radio wavelengths are typically larger than the structures visible in optical wavelengths. Radio galaxies are a class of active galactic nuclei (AGN) and are powered by supermassive black holes at the centres of galaxies. The extended emission is produced by synchrotron radiation of highly relativistic particles accelerated by the AGN. Studying radio galaxies helps to understand the effects of massive black holes on their environment (see e.g. \citet{McNamara_Nulsen_2007}). The jets of highly energetic particles emitted by giant radio galaxies potentially play a major role in the creation of cosmic magnetic fields \citep{Vazza_2022}. 

A lot of new radio sources will be discovered with the new generation of radio telescopes (e.g. LOFAR, MeerKAT, and in the future the SKA \citep{LOFAR, MeerKAT, Carilli_2004}). Processing the incoming data is one of the biggest challenges in radio astronomy. The cause is not only the enormous amount of data, but also the higher source density due to the improved sensitivity of the instruments. Novel techniques are required for this purpose. For instance, the SKA data challenges have demonstrated the difficulties of source finding for SKA data \citep{Bonaldi_2021}. Deep learning has been used to automate processes in radio astronomical data reduction, for example in the automatic flagging of data (see e.g. \citet{Mosiane_2017}). Another example is the work by \citet{Mesarcik_2020} who have used a Variational Autoencoder (VAE) in combination with other methods to automatically inspect data to diagnose system health for modern radio telescopes. Commonly large amounts of labelled training data are required for supervised algorithms, which are not always available. 

Morphological classification of radio sources can be achieved by deep learning models trained on well-understood data sets. \citet{Aniyan_2017,Alhassan_2018,Tang_2019,Samudre_2021, Maslej_2021} use \glspl{cnn} trained on data from the FIRST (Faint Images of the Radio Sky at Twenty-Centimeters) survey \citep{FIRST_1995} for the classification of radio galaxies. The architectures of the neural networks for classification are inspired by the AlexNet \citep{alexnet}. For approaches in radio galaxy classification that use non-standard \glspl{cnn} and other techniques , see e.g. \citet{Lukic_2019,Bowles_2020,Scaife_2021,sadeghi_morphological-based_2021,Ma_2019, Wu_2019, Ntwaetsile_2021}.

In other areas of astronomy, similar morphological classification problems arise, e.g. for classification of optical galaxies \citep{Lintott_2008,Nair_2010}, and of gravitational lenses \citep{Petrillo_2017}. Here supervised methods of machine learning have been applied with some success, see e.g. \citet{10.1093/mnras/staa501, 2021A&A...648A.122V, Walmsley_2019, Huertas_Company_2023}.

However, the existing number of radio sources with morphological labels is limited (the MiraBest data set contains 1254 FRI, FRII and hybrid FR sources \citep{Porter_zenodo_2020}). These class labels are typically extracted from catalogues created and curated manually by experts.
Small data sets used in the training of deep learning models for galaxy classification can be enlarged by data augmentation \citep{Maslej_2021}, e.g. by applying random rotations and reflections to the images (classical augmentation). A different approach based on equivariance implements the symmetry constraints of the problem directly in the construction of the model \citep{Bowles_2021,Scaife_2021}. This may help classifiers to understand symmetries without relying exclusively on augmentation and may be particularly useful for problems with sparse data. 

\begin{figure}
    \centering
    \includegraphics[width=.45\textwidth]{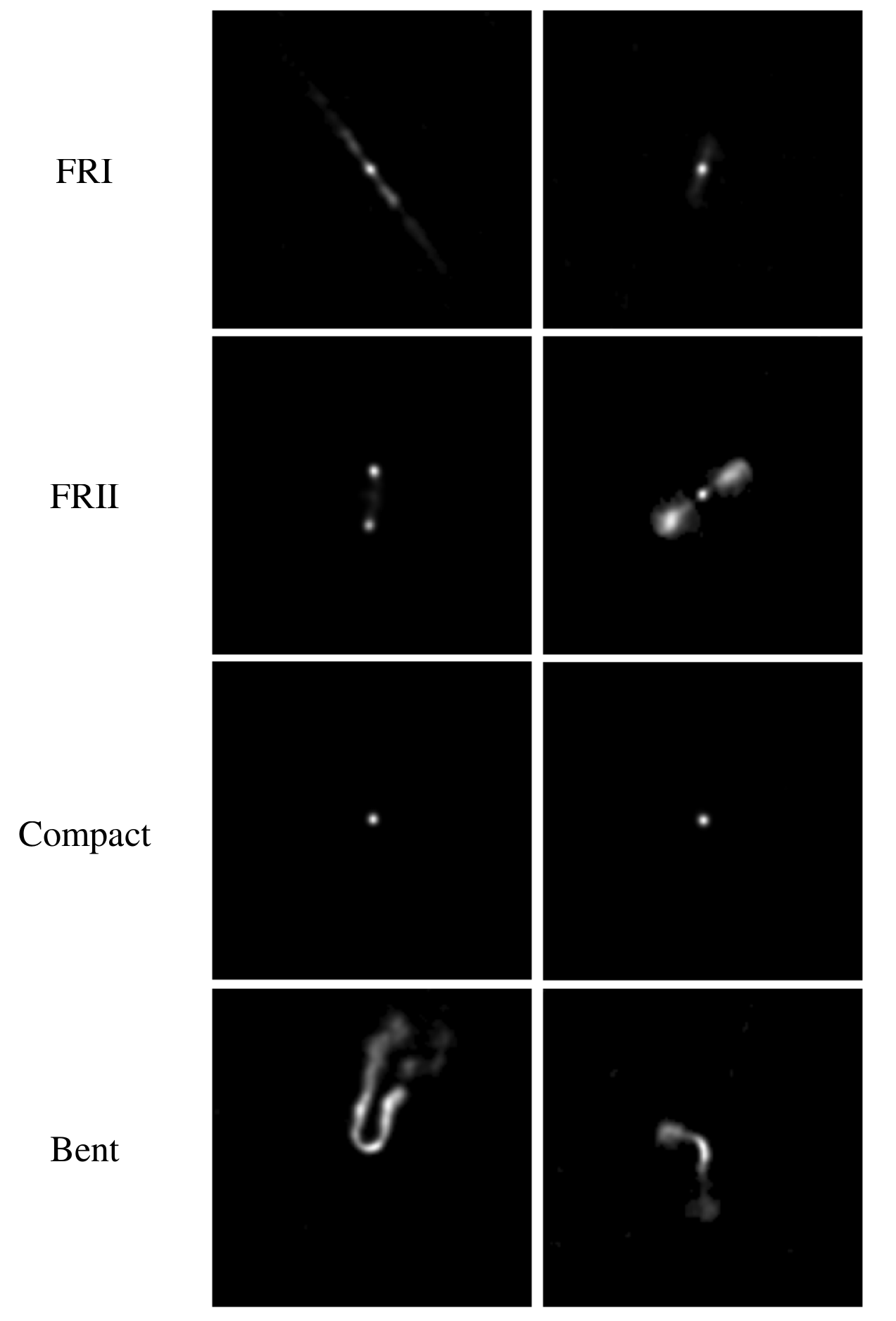}
    \caption{Class definition for FRI, FRII, Compact and Bent. For the Bent class we show an example of a NAT source in the left and a WAT source in the right panel.}
    \label{fig:my_label}
\end{figure}

In this work, we investigate a novel application of generative models to enhance the available training sets. For this augmentation technique, multiple neural networks are combined to learn the underlying distribution of a data set. We focus on the task of classifying different morphological types of radio galaxies. The morphological classification scheme by Fanaroff-Riley is fundamental for such applications \citep{FR_1974}. For the class FRI, the unique maximum of the radio emission resides in the centre of the source and the surface brightness decreases along the jets. For FRII sources, the two maxima of the radio emissions are located at the edges of the jets and the surface brightness in the centre is lower. As radio sources have a large variety of structures, we consider two more classes. Unresolved and point sources are contained in the Compact class. The Bent class consists of sources for which the angle between the jets differs significantly from 180 degrees. The two sub-types Narrow-Angle Tail (NAT) and Wide-Angle Tail (WAT) are further discriminated by the angle, but are fully subsumed in the Bent class for this study. As in \citet{Alhassan_2018,Samudre_2021}, we study a four-class classification problem, including bent-tail and compact sources in addition to the classes FRI and FRII of \citet{FR_1974}. \autoref{fig:my_label} illustrates the considered classes (FRI, FRII, Compact, and Bent).

Other studies probe the use of generative models to create images of radio galaxies \citep{Ma_8451231,Ma_9023752,Bastien_2021}. These studies are based on VAEs. Generative adversarial networks (GANs) have been applied  to astrophysical images in \citet{Schawinski_2017}. For a semi-supervised GAN application to radio pulsars see \cite{radiopular_GAN}. In \citet{Hackstein_2023} various evaluation metrics are used to compare different generative models trained on optical galaxy images. 

In this study, we investigate whether different radio galaxy classifiers can be improved when training is supported by providing additional data generated with a \gls{wgan}. For similar approaches from different fields see, for example, \cite{Frid-Adar_2018,Zhu_2017,Gowal_2021}. We extend our framework presented in \cite{Kummer_2022} to handle larger ratios between real and generated images. Additional images are only generated when they are needed during training. As before, we start with a simple model, namely a \gls{fcn}. In addition, we apply \gls{wgan}-supported augmentation to a \gls{cnn} and a \gls{vit}, see \citet{ViT}.\\

The long-term goal is to use classification models to process incoming data from new radio telescopes.
For this purpose, classification models need to generalise particularly well. A common problem in astronomy is the scarcity of labelled data in the face of large amounts of new data to process. This is a very different situation as for instance in particle physics, where simulations are highly fine-tuned and experiments are constantly repeated.
In particular, for forthcoming radio surveys, and even for the majority of sources in FIRST, no morphological labels are available. As a result unsupervised, semi-supervised and self-supervised methods have gained attention without reaching the performance of supervised methods \citep{Mostert_2021,Slijepcevic_2022,Slijepcevic_2022_SSL}. The current classification scheme of radio galaxies and our physical interpretation will be challenged by new radio surveys. For instance, \citet{Mingo_2019} detected a large population of low-luminosity FRII sources in the LOFAR Two-Metre Sky Survey (LoTSS), see \citet{Shimwell_2019,LoTSS_2022}, that is not expected from the conventional FR distinction based on radio luminosity. Discoveries of rare morphologies can help to extend our understanding of radio sources, but are potentially prohibited by supervised learning techniques. Unsupervised methods such as self-organising maps can be used efficiently to discover such rare morphologies \citep{Mostert_2021}.

This paper is organised as follows: In \autoref{Sec:Data}, we introduce the data set used for training, validation, and testing. The generative model and its implementation are described in \autoref{sec:wGAN}. The training procedure and the assessment of image quality are discussed in \autoref{sec:GenImg}. The results of the comparison between only classical and classical plus \gls{wgan}-supported augmentation for different classifiers are presented in \autoref{sec:wGANsupAug} before we conclude in \autoref{Sec:Disc}.


\section{Data}
\label{Sec:Data}

\begin{table}
    \centering
    \caption{Number of radio galaxy images per class in the train, validation and test data sets.}
    \label{tab:dataset}
    \begin{tabular}{p{.125\textwidth}rrrrr}
         \toprule
          & \text{FRI} & \text{FRII} & \text{Compact} & \text{Bent} & \text{Total}  \\ \midrule
         \text{5-fold cross train} & 316 & 659 & 232 & 198 & 1405 \\ 
         \text{5-fold cross valid} & 79 & 165 & 59 & 50 & 353 \\ 
         \text{test} & 100 & 100 & 100 & 100 & 400 \\ 
         \midrule
         \text{total} & 495 & 924 & 391 & 348 & 2158 \\
         \text{relative frequency}  & 0.23 & 0.43 & 0.18 & 0.16 & 1 \\
         \text{in total} & & & & & \\
         \bottomrule
    \end{tabular}
    
\end{table}

We combine different catalogues \citep{Gendre_2008, Gendre_2010, Miraghaei_2017, Capetti_2017b,Capetti_2017a, Baldi_2017, Proctor_2011} that characterise radio sources from the FIRST survey to create a data set of 2158 radio galaxy images with morphological labels. The labelling in the catalogues is typically performed by experts by considering radio images and the corresponding optical counterparts. We group radio sources into 4 classes, namely FRI, FRII, Compact and Bent. The source coordinates are compared between catalogues to remove duplicates. 
 Sources that appear with different labels are regarded as ambiguous and are removed entirely. More details on the acquisition of the data set can be found in \citet{GRIESE2023108974}. The data set is published on zenodo \citep{Griese_zenodo_2022} and on GitHub (\url{https://github.com/floriangriese/RadioGalaxyDataset}). 
The radio galaxy images of the FIRST survey are collected from the virtual observatory skyview\footnote{\url{https://skyview.gsfc.nasa.gov}}. We start from the original images with a size of (300$\ \times \ $300) pixels. Then we adopt the preprocessing procedure from \citet{Aniyan_2017}. In particular, we set all pixel values below three times the local RMS noise to the value of this threshold. We apply classical augmentation to all images during training consisting of random rotations and reflections of the base image. This augmentation is done every time an image comes up in the training loop, so that the augmentation factor simply depends on the number of iterations of the training procedure. Consequently, classical augmentation retains the class imbalance present in the base image set. The augmented images are then cropped to the input size of our generative network, i.e. to (128$\ \times \ $128) pixels. Subsequently, the pixel values are rescaled to the range [-1, 1] to represent floating point greyscale images.

We separated 100 sources per class from the data set for the final evaluation of our models. For validation purposes during training (e.g. choosing the best model), we use a 5-fold cross-validation. Therefore, we do not need a separate validation set. As a result, we lose less training data. In particular, we split the training set into five blocks and did five separate training runs. For each of these runs one of the five blocks was used as the validation set and the remaining four blocks represented the corresponding training set. The quantities per class and per split are shown in \autoref{tab:dataset}.

%

\section{Wasserstein GAN}
\label{sec:wGAN}
The ability to learn representations of underlying statistical distributions of data sets makes generative models a powerful tool for the creation of additional data points. In particular, sampling from those representations allows to speed up conventional simulation techniques significantly and may be useful for further subsequent treatments \citep{Buhmann_2021,Hadrons_2021}.

Three different categories of generative models are well-established: GANs, VAEs, and flow-based models. Diffusion models represent a relatively new development in this area. In this work, we focus on GANs. They consist of two neural networks: a generator $G$ that generates fake images from a noise vector $Z$ and the discriminator $D$ that discriminates between real and fake images. This architecture was first introduced in \cite{goodfellow2014generative,salimans2016improved}. In a two-player minimax game, the generator learns to create fake images, which become less and less distinguishable from the real ones in the course of the training. The loss function for this setup reads~\citep{goodfellow2014generative,salimans2016improved}:
\begin{equation}
    L=\min_G \max_D \mathbb{E}[ \log D(x) ] + \mathbb{E} [\log (1-D(G(z)))],
\end{equation}
where $x$ represent real samples and $G(z)=\tilde{x}$ generated samples.

For this project, we employ a variant of the standard GAN setup called \gls{wgan} that uses the Wasserstein-1 metric, also referred to as the Earth Mover's distance, as main term in the loss function ~\citep{arjovsky2017wasserstein}. This loss function is calculated as
\begin{equation}
    L=\sup_{f\in\text{Lip}_1}\{\mathbb{E}[f(x)-\mathbb{E}[f(\tilde{x})]\}, 
\end{equation} 
where $f$ denotes a 1-Lipschitz function that
is learned during the training procedure.
The discriminator of a standard GAN is transformed into a critic and is used to estimate the Wasserstein
distance between real and generated images. Hence, the absolute value of the loss function is correlated with the image quality, resulting in the name change.
Additionally, the training of \glspl{wgan} is often more stable and more likely to converge than standard GAN setups. To approximate the Wasserstein-1 metric by use of a critic network, it has to be ensured that the 1-Lipschitz constraint is fulfilled. This is achieved by applying a gradient penalty term to the loss function as in~\cite{gulrajani2017improved} 
\begin{equation}
    L=\lambda\ \mathbb{E}[(\|\nabla_{\hat{x}}f(\hat{x})\|_2 -1)^2]
\end{equation} 
for random samples $\hat{x}\sim\mathbb{P}_{\hat{x}}$.

Since we work with image data, it has proven to be the most promising approach to construct a \gls{wgan} setup based on convolutional layers \citep{Radford_2015}. The generator receives a noise tensor of size 100$\times$1 and a class label $y$ and, through multiple layers of 2D transposed convolution operators, enlarges this to a 128$\ \times \ $128 tensor, consistent with the dimensions of real images. The critic is given either real or generated images, as well as the class label $y$. 
The output of the critic is a single real value, which represents the belief of the critic for the image to be real. Generator and critic are trained intermittently, where the critic has five training cycles per training cycle of the generator. When training the generative model, it is necessary to apply classical augmentation such that the symmetries of the training set are also present in the generated data sets, and to avoid introducing a bias due to the limited number of training examples.\\

\begin{figure}
    \centering
    \includegraphics[width=.48\textwidth]{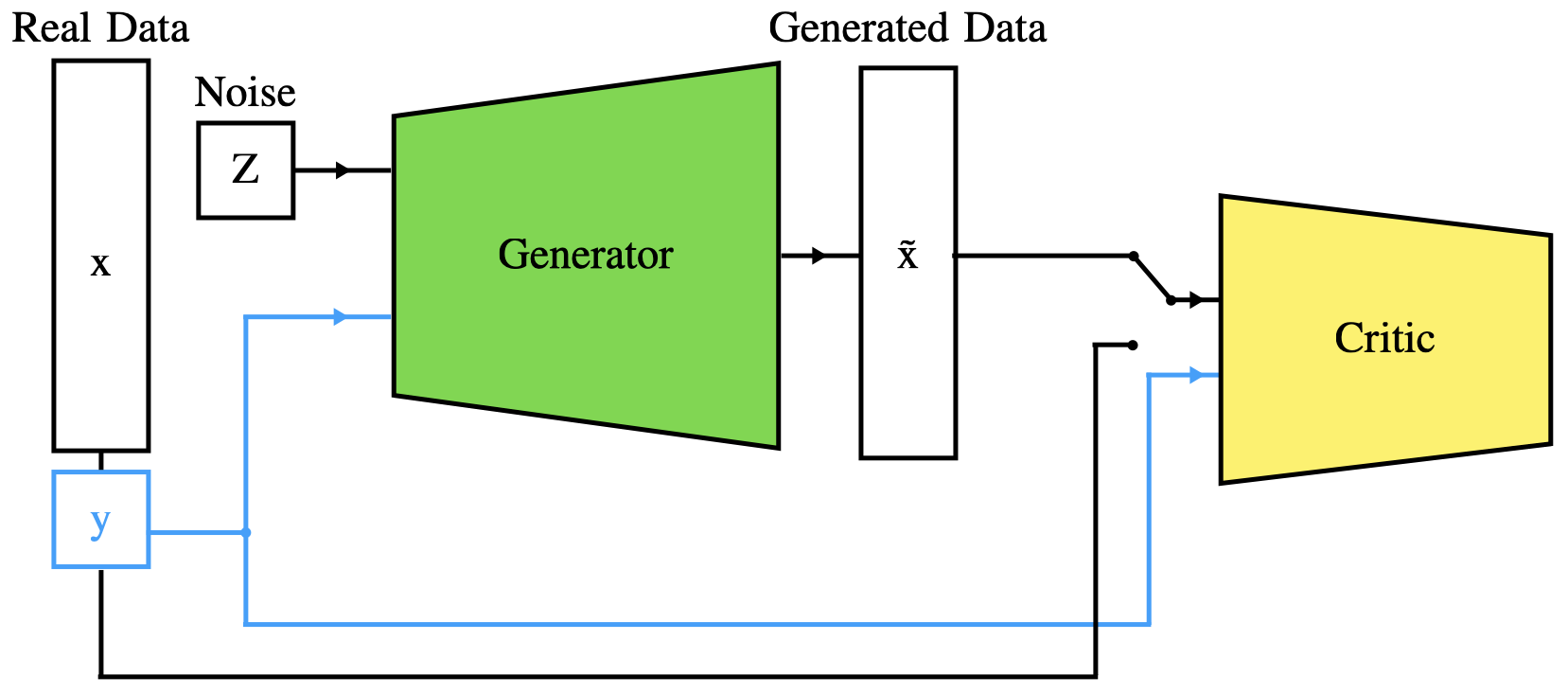}
    \caption{Schematic of the \gls{wgan} architecture, where $y$ denotes the class label of real $x$ or generated images $\tilde{x}$}
    \label{fig:wgan_schematic}
\end{figure}

Morphologies of radio galaxies are diverse and result in very different images. Consequently, it is reasonable to condition the networks with the class label $y$ such that a combination of image and class label is provided to the networks. In particular, this allows applying supervised learning techniques on the output of the generator. For our setup, this is achieved for the generator by applying a 2D transposed convolution operator on a matrix of image dimensions filled with the class label. The transpose-convoluted layer is then concatenated to the first transpose-convoluted layer of the noise tensor. Batch normalisation in 2D and ReLU (Rectified Linear Unit) activation functions are used. The concatenated tensor is then passed through five additional 2D transposed convolutions, where no normalisation or activation is applied after the last layer. Instead, the individual pixel values are clipped to [-1,1] for conversion to grayscale.
The critic is built analogously, but uses 2D convolutional layers, resulting in a single output node representing the critic score for image quality. Here, layer normalisation and Leaky ReLU functions are used except for the last layer. The Leaky ReLU activation function is an attempt to avoid the "dead Neuron" phenomenon of the pure ReLU function, where any gradient information is lost if the input is negative. This makes the critic more stable against sub-optimal starting points. The Layer Norm computes the normalisation over the features instead of batches. A schematic of the wGAN setup can be found in \autoref{fig:wgan_schematic}. For more details on the architectures see \autoref{tab:wgan_critic} and \autoref{tab:wgan_generator}.

\section{Results of image generation with a wGAN}
\label{sec:GenImg}
\subsection{Training}
For each choice of training and validation data in the cross-validation procedure a \gls{wgan} training run is launched on the corresponding training set. The training is performed with a single NVIDIA A100 GPU provided by the Maxwell cluster at DESY for 40k generator iterations, i.e. weight updates. A batch size of 400 is chosen and one training run takes roughly seven hours to complete. The choice of the batch size did not have a strong impact on the performance of the model, so that we chose a size that still comfortably fits into the GPU's memory, while being large enough to fully profit from the computing speed-up of larger batches. The generator and critic weights are saved every 250 iterations, allowing to scan for the best training state later on, as described in the following section. Choosing such an iteration for every model and training run is necessary as \gls{wgan} training runs generally do not converge fully but rather fluctuate around an optimal value. This means that it is not instructive to simply use the final state of the model after training and instead other metrics need to be studied to choose an optimal working point. While comparing different model setups, we are only interested in the performance of these optimal working points. All models are implemented and trained in PyTorch \citep{pytorch}. For an overview of training details we refer to \autoref{tab:hyperparam}. The choice of hyperparameters is inspired by values obtained by \citet{Buhmann_2021}. With the exception of the learning rate, other hyperparameters have not been further optimised.

\subsection{Evaluation of image quality}
In this section, we present images created using the generator of the \gls{wgan} and examine the quality of the generated images in several ways.

\subsubsection{Distribution-based comparison}

We define a set of distributions to compare generated images with the training data set, in order to determine the quality of generated images and thus to find the best performing training iteration. This includes normalised histograms of pixel intensities, the number of pixels with an intensity greater than zero and of the sum of intensities. These histograms are compared for each class individually and the relative mean absolute error (RMAE) between the generated set of 10k images and the training set is computed. The RMAEs for the different distributions are summed up to yield a single figure-of-merit (FOM), where the \gls{wgan} training iteration with the lowest FOM value is used in the following as the best model. 
This procedure is followed for each of the four classes separately, i.e. we allow a different iteration of the generator training to yield the best model for each class. The chosen distributions are commonly used for images (e.g. photography), but it is important to note that they do not specifically contain information on the shape of the radio galaxies within these images. The choice of RMAE is based on its very fast computing time and robustness against empty bins while we acknowledge that other test metrics can be used. \\

\begin{figure}
\begin{minipage}{.5\textwidth}
    \includegraphics[width=1\textwidth]{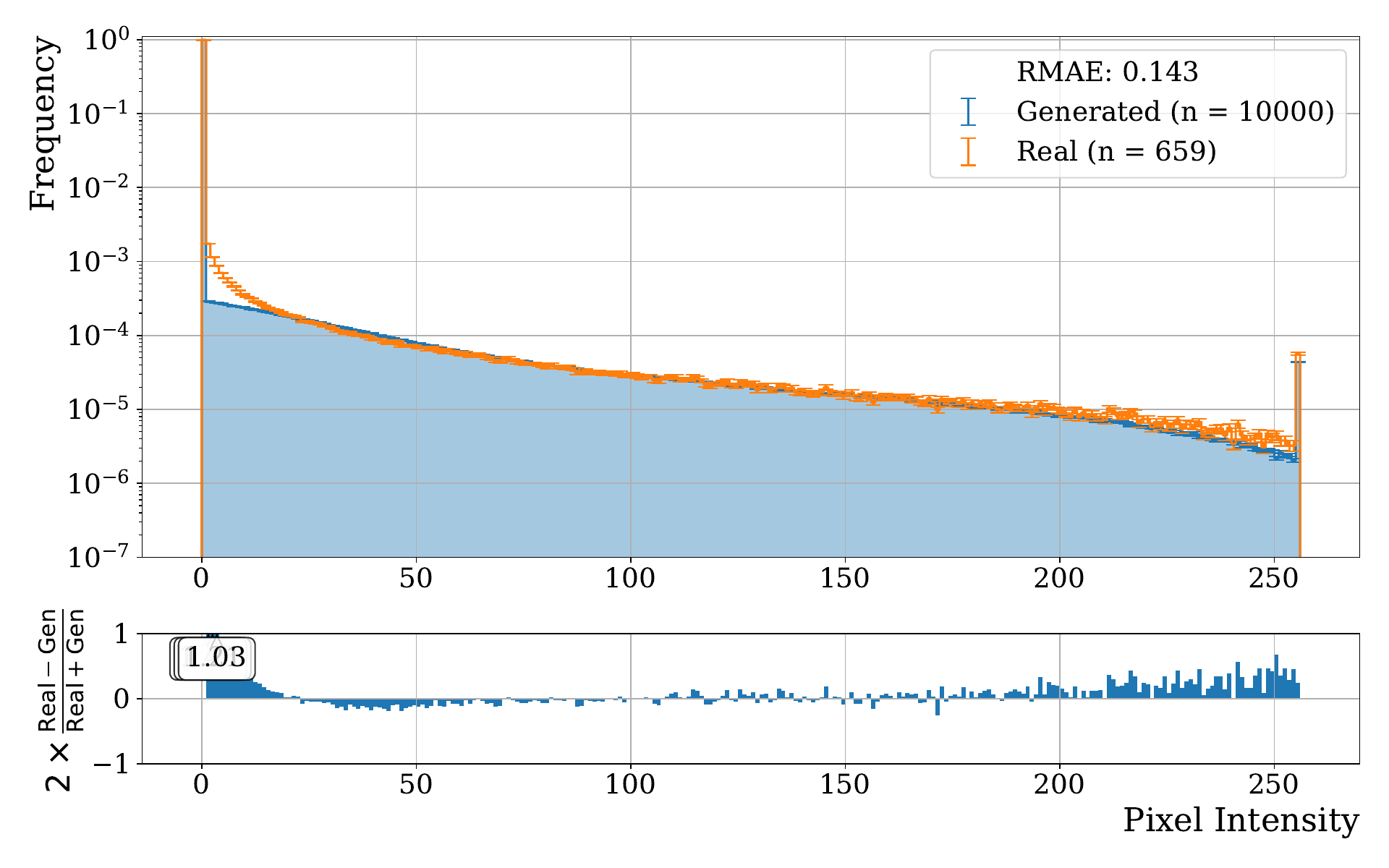}
    \subcaption{Pixel intensities of FRII sources.}
    \label{fig:histogram_int}
\end{minipage}
\begin{minipage}{.5\textwidth}
    \includegraphics[width=1\textwidth]{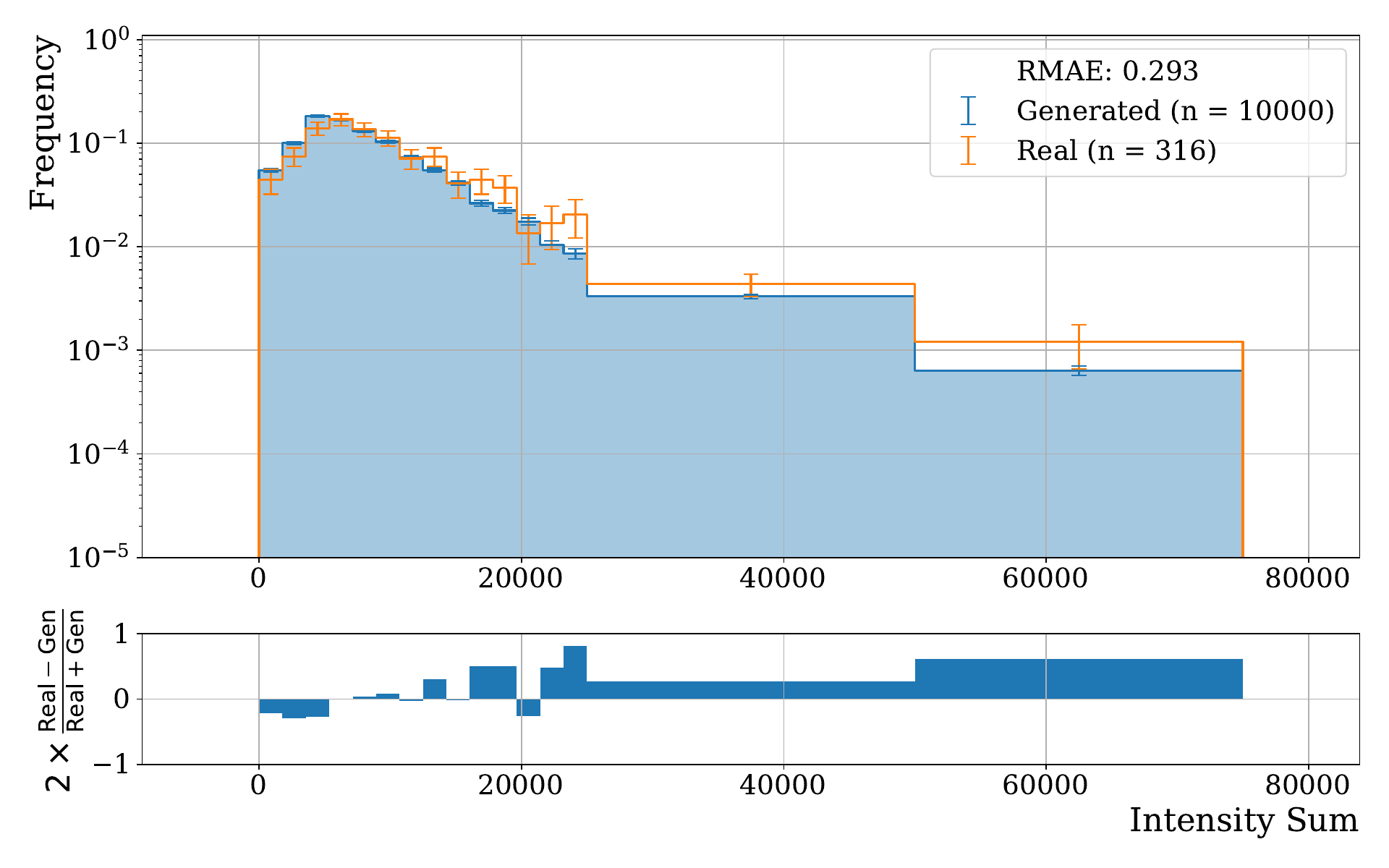}
    \subcaption{Sum of pixel intensities of FRI sources.}
    \label{fig:histogram_sumi}
\end{minipage}
\begin{minipage}{.5\textwidth}
    \includegraphics[width=1\textwidth]{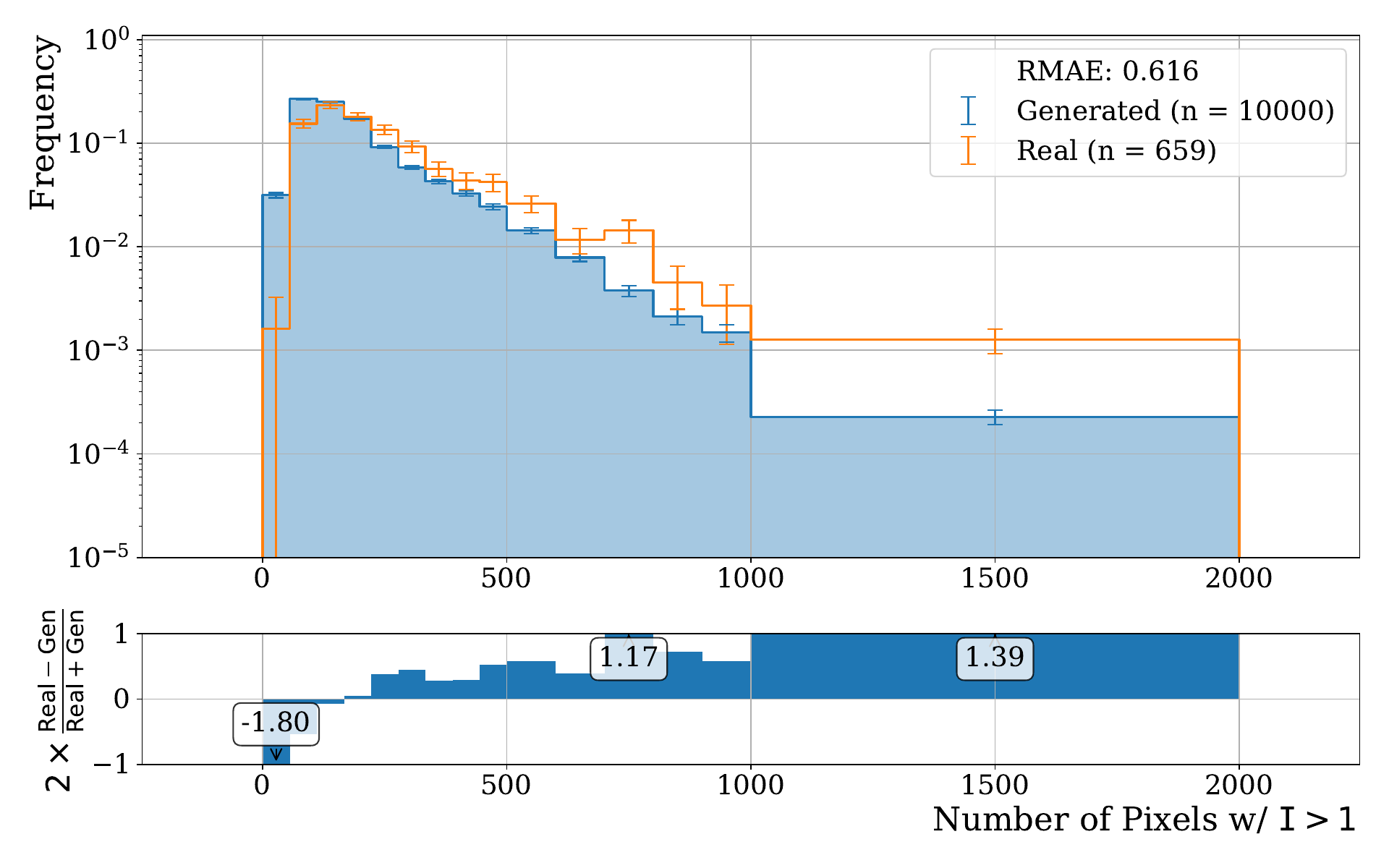}
    \subcaption{Number of activated pixels of FRII sources.}
    \label{fig:histogram_activated}
\end{minipage}
\caption{Examples of image quality measures comparing histograms of real (orange) and generated (blue) images for the generator training iteration with the lowest combined RMAE for the corresponding class. The per-bin relative error is shown in the bottom of each panel. Histograms shown here are chosen arbitrarily from the first cross-validation fold.}
\label{Fig:wGAN_metrics}
\end{figure}

Arbitrarily chosen examples of these distributions are shown in \autoref{Fig:wGAN_metrics}, where the distribution of the real images is shown in orange and the distribution of the generated images in blue. The uncertainty for each bin is given by the square-root of entries in that bin before normalisation. The bottom panels in this figure show the per-bin divergence between the distributions, where absolute deviations larger than 1 are indicated by the corresponding value written in boxes. Here, only examples from the first cross-validation fold (of five) are shown.\\

Overall, the distributions of the generated images tend to follow the distribution of the real images. Nevertheless, the generated images have difficulties in recreating very low, but non-zero, intensities. This can be seen for pixel values between 1 and 20 in \autoref{fig:histogram_int}, which directly translates into under-representing the number of pixels with an intensity $I>0$ in \autoref{fig:histogram_activated}.

\begin{figure*}
    \centering
    \includegraphics[width=1.0\textwidth,trim={3cm 4cm 3cm 4cm},clip]{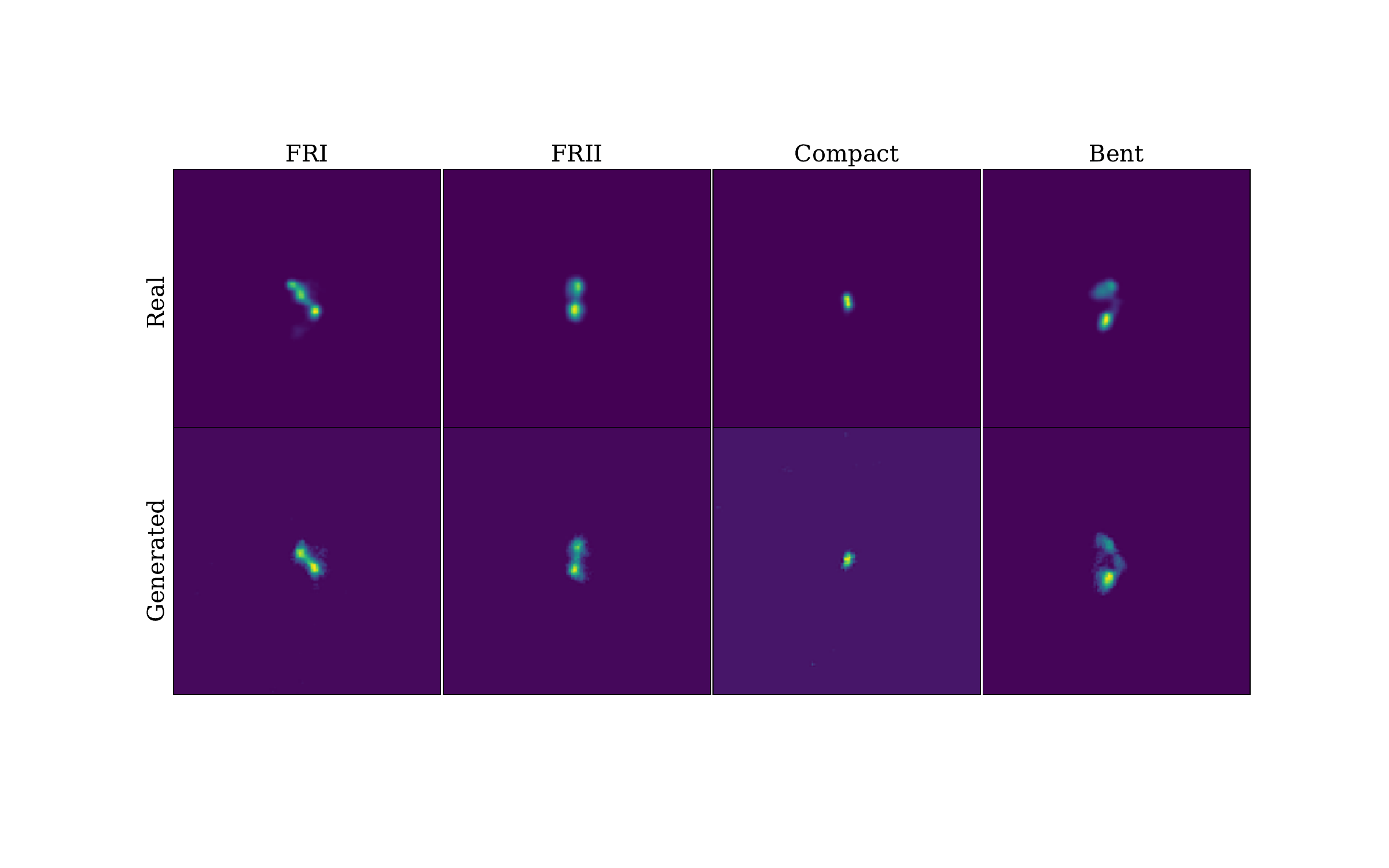}
    \caption{Closest matching pairs in terms of a pixelwise difference between generated and real images for each class. The set of real images is the full training data set over all cross-validation folds; the generated data set consists of 5k images per class. Images are aligned according to the first principal component.}
    \label{fig:image_examples}
\end{figure*}

\subsubsection{Visual comparison}

In order to get a visual idea of image quality, we generated a set of 5k images per class and compared them to the full training data set over all cross-validation folds. 
The images are rotated so that their principal components are aligned. Subsequently, we compute the pixel-by-pixel difference for all possible pairs of real and generated images. All classes also include a few difficult to define sources with rather small spatial extension that are easy to emulate but do not show the generator's capability of reproducing the more interesting extended sources. Thus, we only consider images with an intensity sum of at least 15k (5k) for the extended (compact) radio galaxies. We show the resulting closest pairs for each class in \autoref{fig:image_examples}. By eye, the generated images appear very similar to the analogue real images, indicating a good performance of the generator setup in terms of fidelity. In addition, the diversity of the generated data is crucial for the study in \autoref{sec:wGANsupAug}. To also get an impression of this diversity we show a random set of generated images in \cref{App:img}.

\subsubsection{Classifier-based comparison}

\begin{figure}
\begin{minipage}{.48\textwidth}
    \includegraphics[width=\textwidth]{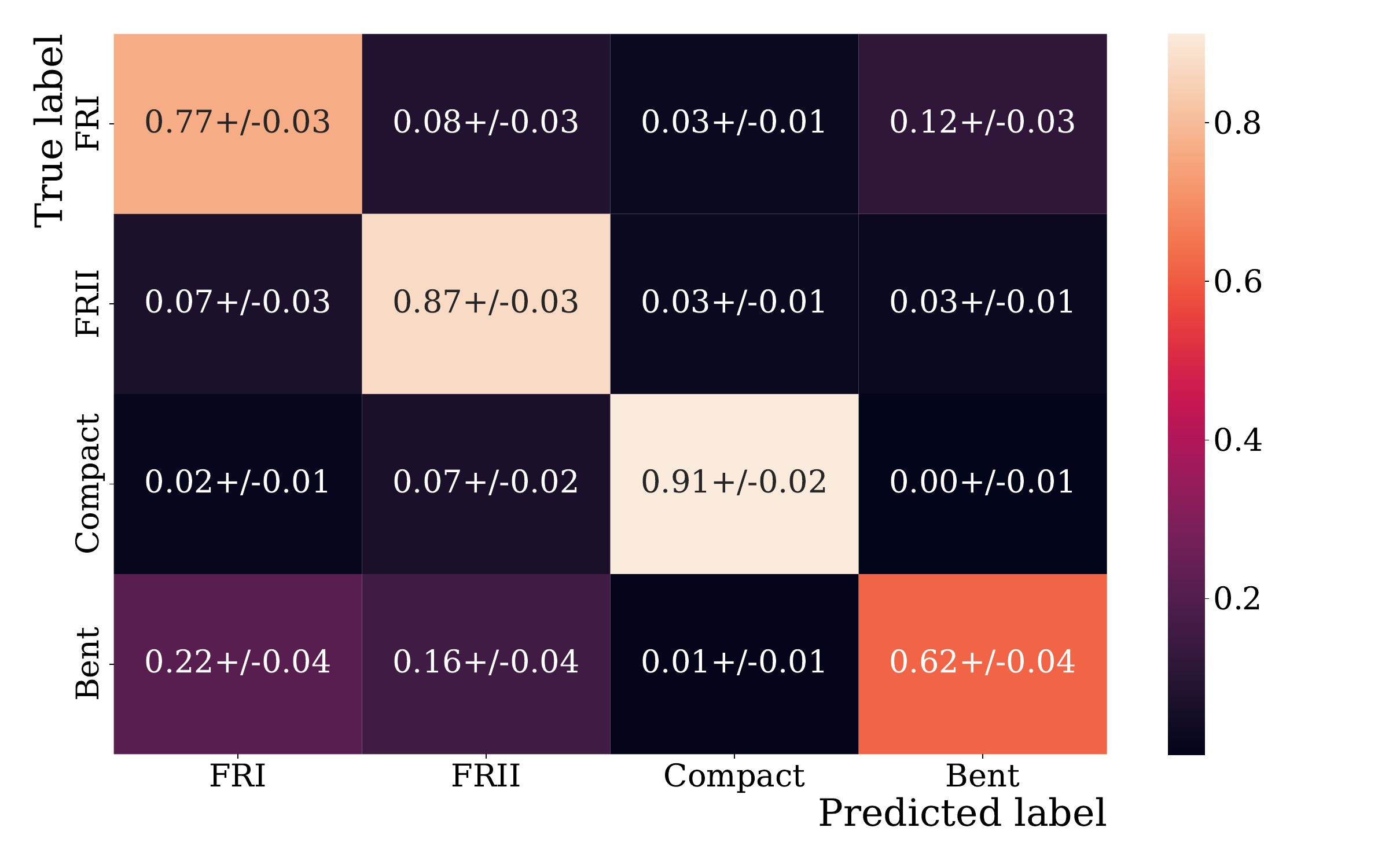}
    \subcaption{Real test sample.}
\end{minipage} \\ 
\begin{minipage}{.48\textwidth}
    \includegraphics[width=\textwidth]{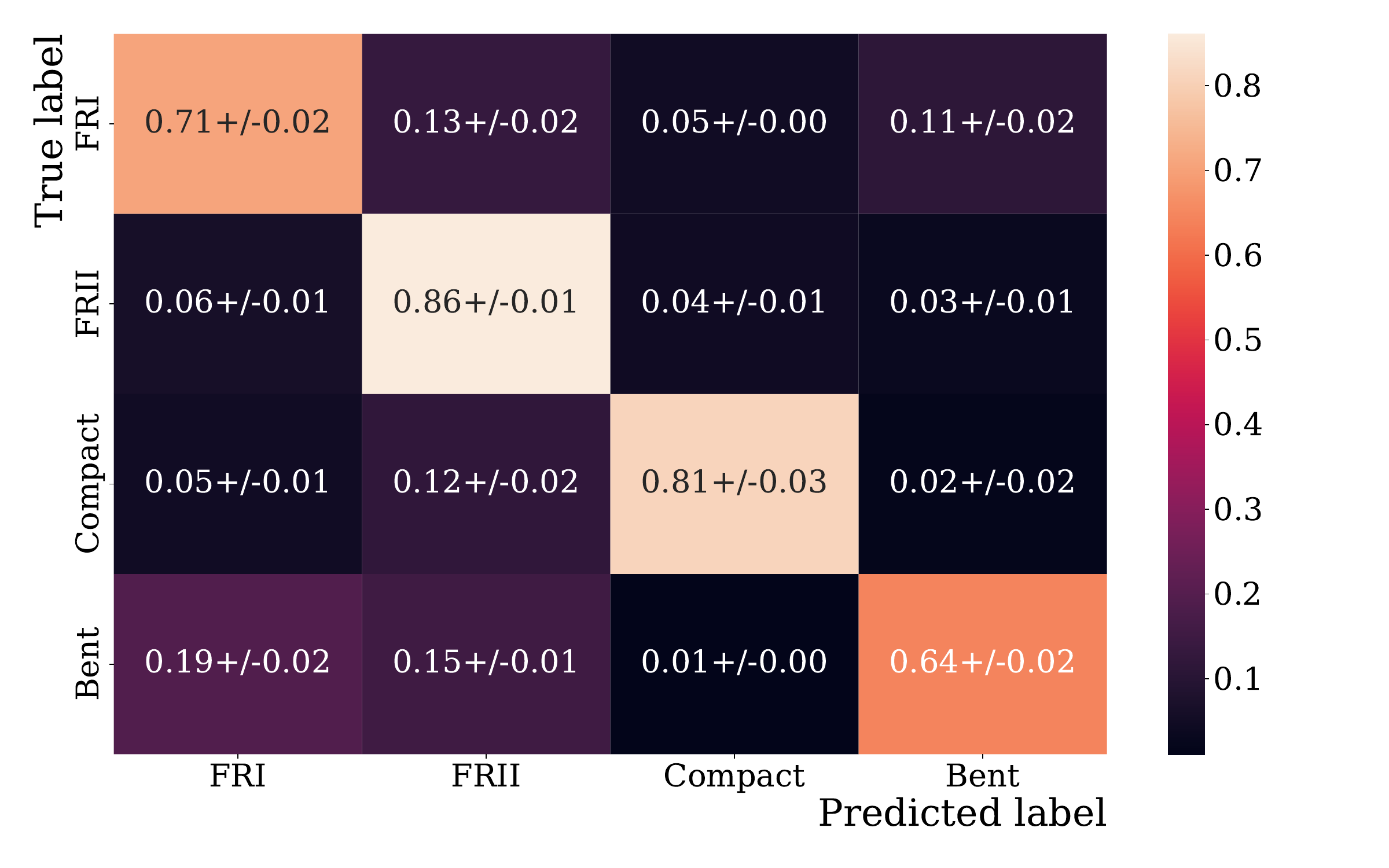}
    \subcaption{Generated sample.}
\end{minipage}
    \caption{Confusion matrices on the real only test data set (above) and a data set of 4k generated images, where each of the generators pertaining to a cross-validation fold contributes 200 images per class (5 generators $\times$ 4 classes $\times$ 200 images). Matrices are row-wise normalised and both data sets are class balanced. The values represent the mean and standard deviation of the confusion matrices obtained from each of the classifiers trained on the cross-validation folds with classical augmentation only.}
    \label{fig:confusion_example}
\end{figure}

Next, we use a CNN trained solely on the data set of real images to assess the image quality further.
We compare the performance of the same classifier evaluated on the real test set and a set of generated images. The architecture of the CNN used for this experiment is summarised in \autoref{tab:cnn_classifier} and the hyperparameters in \autoref{tab:hyperparam}. A comparison of the confusion matrices on both sets tests for any bias introduced by the image generation. In particular, we evaluate the conditioning on the class labels. In the top panel of \autoref{fig:confusion_example}, we show the confusion matrix of the classifier on the real test set. Comparing this to the confusion matrix obtained by the same classifier on a set of generated images on the bottom panel of \autoref{fig:confusion_example}, we find that the class conditioning of the generated images works overall quite well. However, confusion for images of the class FRI with the predicted classes FRII is enhanced on the generated test set. The classification performance of the Compact class is decreased on the generated test set, where particularly the misidentification of true Compact class images as FRII images is increased. This might be due to the fact that some FRII-like sources resemble a combination of two compact sources. Confusion for true Bent class images predicted to be of the FRI class is slightly reduced. The confusion between FRI and bent-tail sources is expected to be large as these classes contain sources that have faint, smeared out radio structures. In contrast, FRII and compact sources typically share sharp margins. 

\section{Results of classifier training using wGAN-supported augmentation}
\label{sec:wGANsupAug}

We assess the new approach of supplementing the training set with generated images by comparing the performance of different classifiers (each trained on different setups with increasing amount of generated data). Our benchmark is the performance of the classifier trained on the original training set. We test the performance of the classifier trained with the original training set plus simulated images by the generator of the \gls{wgan} against this benchmark. We start with a \gls{fcn} (see \autoref{tab:fcn_classifier}). Subsequently, we increase the complexity of the classifier by training a CNN (see \autoref{tab:cnn_classifier}).
Finally, we apply our framework to a state-of-the-art classifier, namely the \gls{vit} \citep{ViT}.  Inspired by the performance of transformers in natural language processing, like BERT \citep{https://doi.org/10.48550/arxiv.1810.04805}  and GPT \citep{radford2018improving, radford2019language, https://doi.org/10.48550/arxiv.2005.14165}, vision transformers are frequently used in computer vision tasks e.g. classification, object detection and segmentation \citep{10.1145/3505244, shamshad2022transformers, ulhaq2022vision}. The self-attention mechanism enables learning long range relationships between items within a sequence. Further, the architecture provides a scalability to high complexity models \citep{10.1145/3505244}. As the transformer assumes less prior knowledge than a \gls{cnn} based model, it requires more training data, Thus, the transformer models are typically pre-trained on large-scale data sets to learn more general representations and afterwards the learned representations are fine-tuned to the task with limited data \citep{10.1145/3505244}. In our case, we use the default ViT-B\_16 vision transformer configuration with pre-trained weights from the ImagetNet21k data set\footnote{For the adopted \gls{vit} implementation see \url{https://github.com/lucidrains/vit-pytorch} and for the corresponding weights see \url{https://github.com/google-research/vision_transformer}} with a resetted head layer. The \gls{wgan}-generated images with pixel sizes 128x128 are zero-padded up to 224$\times$224 pixels to fit the pre-trained model input size. As an attention based model, the ViT splits the image into fixed-size patches processed by the transformer encoder.

We generate images on the fly, i.e. each time a generated image is loaded it is newly generated. The images are generated such that the resulting data set is balanced. As a loss function, cross-entropy loss is implemented, weighted for the imbalanced data only runs. The training setups are not optimised to reach maximal classification accuracies. The goal of this study is only to compare classical augmentation with wGAN-supported augmentation for each of the classifiers. We do not compare performance between the three classifiers in detail either. For further training details see \autoref{tab:hyperparam}.

\begin{figure}
    \centering
    \includegraphics[width=.49\textwidth]{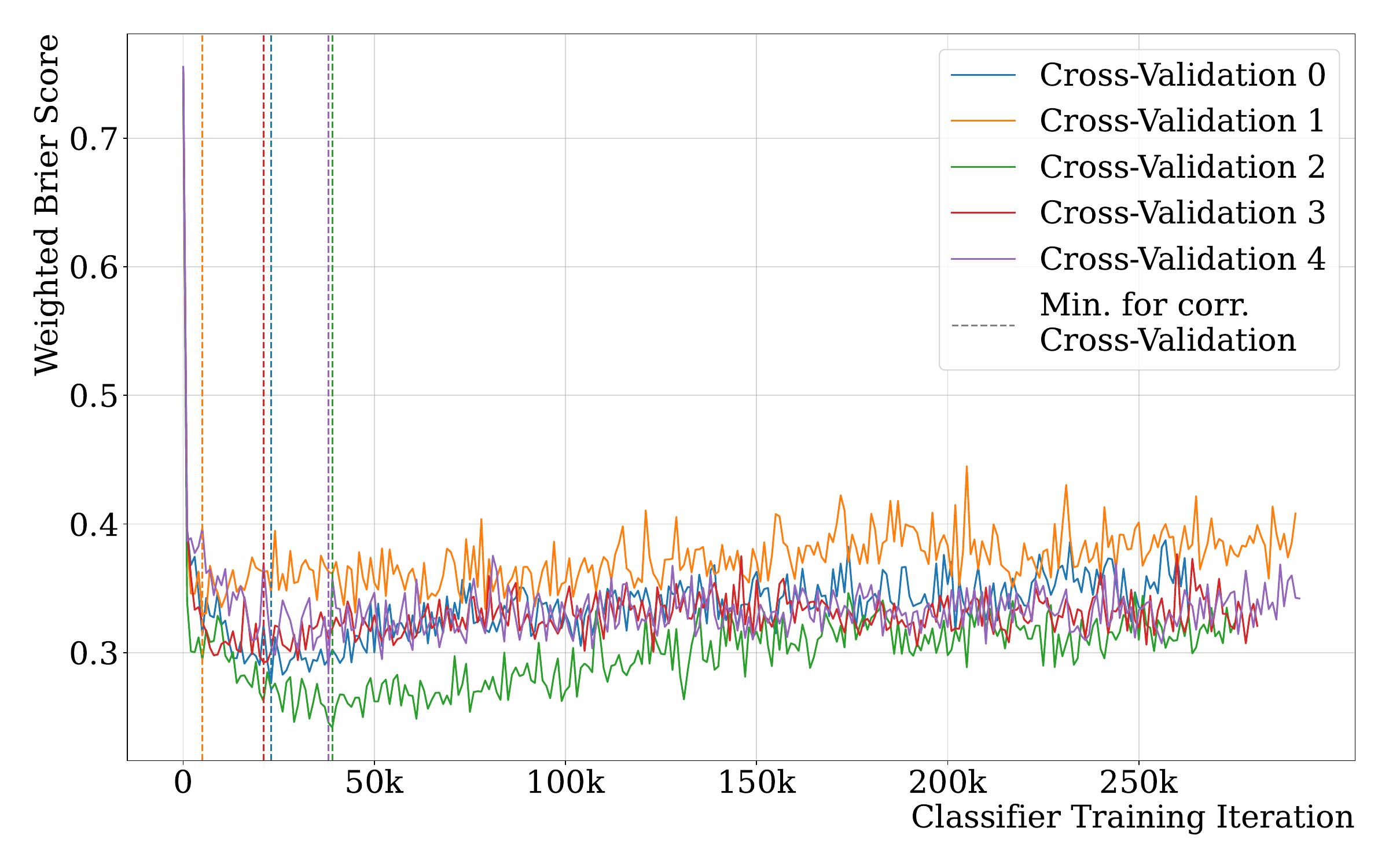}
    \caption{Weighted Brier score on the corresponding cross-validation validation sets per classifier training iteration. The iteration resulting in the minimum score is indicated by a dashed vertical line.}%
    \label{Fig:brier_validation}
\end{figure}

\paragraph*{Evaluation metrics}
To compare the overall performance among different training setups and to determine the best training iteration of a classifier training run (see \autoref{Fig:brier_validation}), we use the multi-class Brier score~\citep{Brier_1950}.
The Brier score is essentially the mean squared error of the predicted probabilities of a classifier for all classes. This has the advantage that also the certainty of the classifier's decision is considered, which winner-takes-all FOMs such as accuracy do not take into account.
For each setup, i.e. for a given ratio between the number of generated $i_g$ and real images $i_r$, denoted $\lambda = i_g/i_r$, we have five models due to the 5-fold cross-validation.

The final evaluation is performed on an independent test set that contains real data only. We use the most commonly applied metric in radio astronomy publications: multi-class accuracy. In order to estimate statistical fluctuations, we average the performance metrics over the five best models of each cross-validation fold.

\begin{figure}
    \centering
\begin{minipage}{.48\textwidth}
    \includegraphics[width=\textwidth]{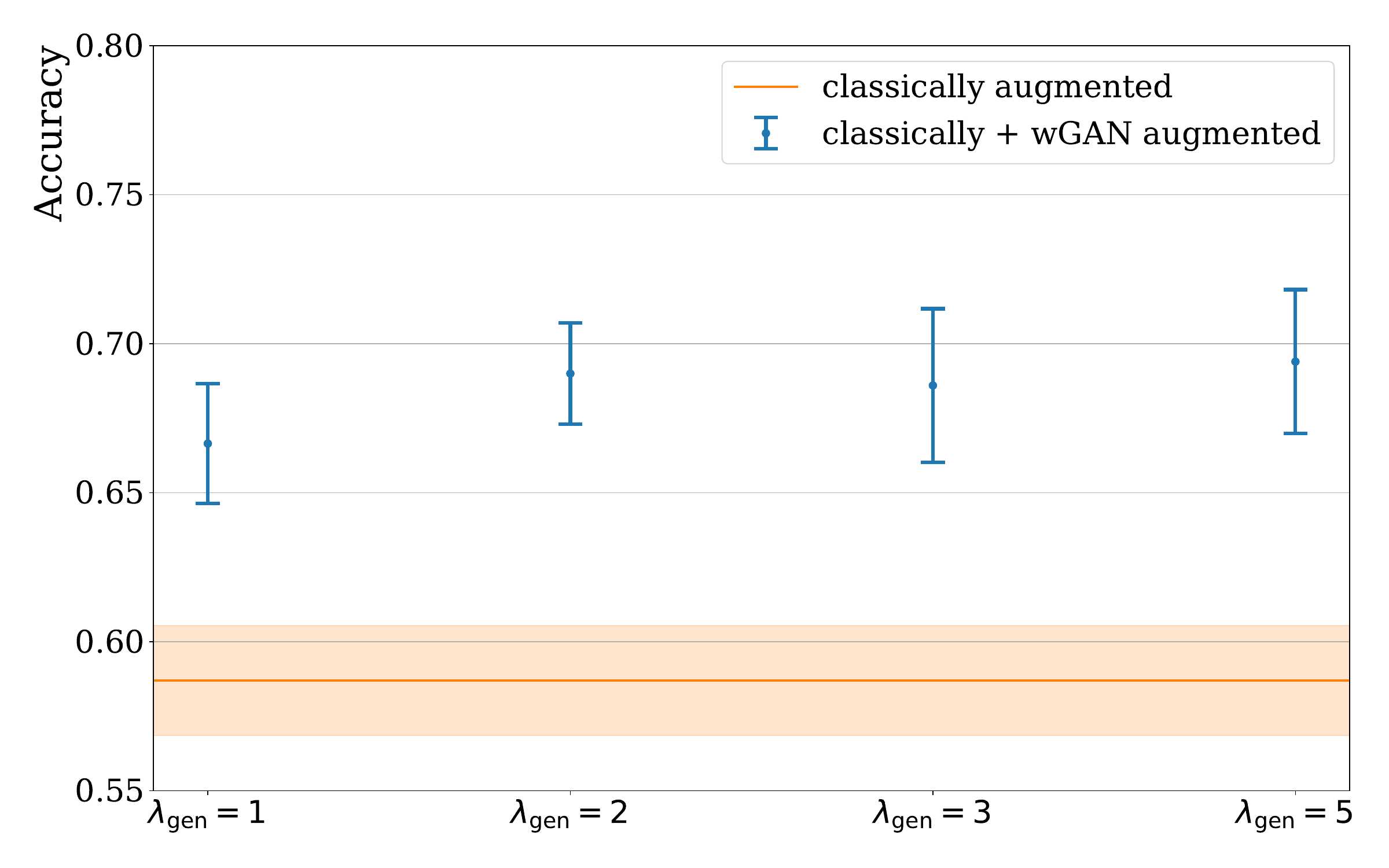}
    \subcaption{FCN}
    \label{Fig:ResultsFCN}
\end{minipage}
\begin{minipage}{.48\textwidth}
    \includegraphics[width=\textwidth]{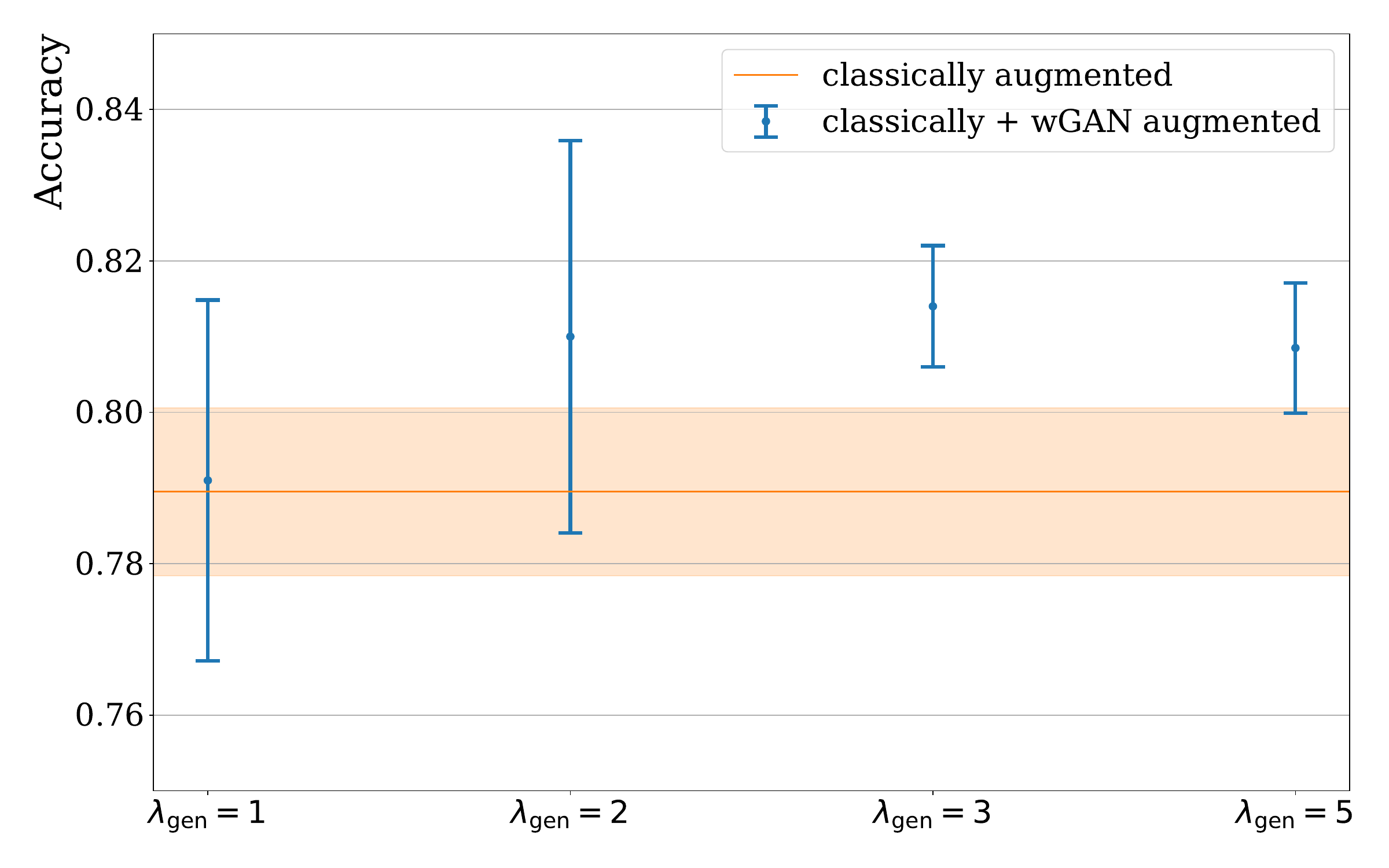}
    \subcaption{CNN}
    \label{Fig:ResultsCNN}
\end{minipage}
\begin{minipage}{.48\textwidth}
    \includegraphics[width=\textwidth]{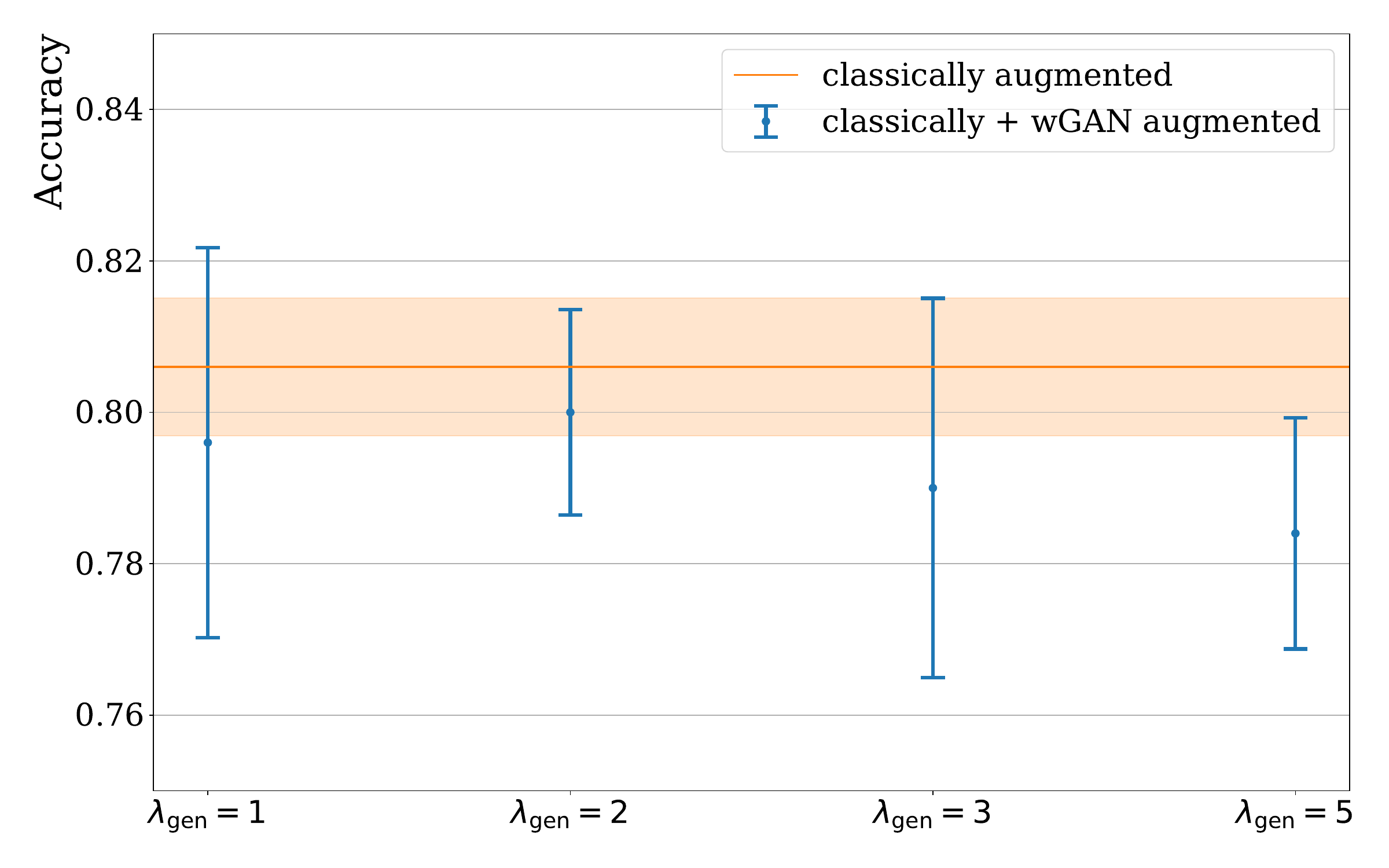}
    \subcaption{ViT}
    \label{Fig:ResultsVIT}
\end{minipage}
    \caption{Accuracy on the test data set for the three different classifier architectures for different training scenarios, where the amount of generated images used to augment the training data set (represented by $\lambda$) is varied. The blue markers (uncertainty bars) represent the mean (standard deviation) of the obtained results over all cross-validation folds for the classically + \gls{wgan} augmented training data sets and the horizontal orange line (area) show the corresponding result for the only classically augmented case.}
    \label{Fig:ClassifierResults}
\end{figure}

\paragraph*{Accuracy} The multi-class accuracy, i.e. number of correct classifications over number of all classifications, on the test data set is shown in \autoref{Fig:ClassifierResults} for the three different classifiers investigated here. The results are shown for different training scenarios, where the number of generated images used to augment the training data set (represented by $\lambda$) is varied. The blue markers (uncertainty bars) represent the mean (standard deviation) of the obtained results over all cross-validation folds for the augmented training data sets and the horizontal orange line (area) show the corresponding result for the real data only case.\\

\autoref{Fig:ResultsFCN} presents the results for the \gls{fcn}, which yields an improvement in accuracy of $\SI{17.5+-4.7}{\percent}$ over the baseline setup at $\lambda=2$. All augmented training setups outperform the real data only case which reaches an accuracy of $\SI{58.7+-1.8}{\percent}$.\\

The highest obtained average for the \gls{cnn} classifier is reached for $\lambda = 3$, as can be seen in \autoref{Fig:ResultsCNN}, which is $\SI{3.0+-1.8}{\percent}$ higher than the real data only baseline at $\SI{78.9+-1.1}{\percent}$. 
The highest obtained average using \gls{wgan} augmented training data for the \gls{vit} classifier is reached at $\lambda = 2$, see \autoref{Fig:ResultsVIT}, which is $\SI{0.7+-2.0}{\percent}$ lower than the real data only baseline at $\SI{80.6+-0.9}{\percent}$.
For additional performance analyses per class we refer to \cref{App:AddPlots}.

\section{Discussion and Conclusion}
\label{Sec:Disc}
The approach used for the study presented here, utilising a \gls{wgan}, is novel to the field of radio astronomy. We are able to generate highly realistic images of radio sources of the four different radio galaxy classes. For this, we rely on the good agreement between the image metric distributions, such as the pixel intensity histogram, between real and generated images, as well as the good agreement between the confusion of a \gls{cnn} classifier trained only on real data obtained on a real data only test set and a generated data only test set. Particularly the latter, provides confidence for the class conditioning of the generator.

Following a visual inspection, we note that the generated images tend to have sharper edges, i.e. low intensity pixels directly next to high intensity pixels. This is not the case for real images, which are smeared due to detector resolution effects. Resolving these issues would yield even more realistic generated images.

However, we do not observe issues known from other state-of-the-art generative networks in radio astronomy. VAE-based models suffer from different noise levels between generated and training data or pseudo-textures and pseudo-structures (see e.g. \citet{Bastien_2021}). The results of this study therefore constitute a major improvement in generated image quality.

This high quality of the images allows us to use them to improve the training of an external classifier, called \gls{wgan}-supported augmentation here. This represents an extension to realistic data of the studies done in \citet{GANplify_2021}, which showed that statistical information contained in a simplistic toy training data set can be augmented using generative models. Another extension of this study to more realistic data in the field of particle physics is given in \cite{Calomplification_2022}.

We find in agreement with these studies that generated images individually contain less information than real data. An additional test presented in \cref{App:AddTest} shows that the performance of a classifier worsens if the amount of real training data is reduced and replaced by generated data. However, the statistical power of the training set can be increased by the inclusion of generated data.

Here, we are able to show that adding generated images to the training data set does clearly improve the classifier performance on a real data only test set for the \gls{fcn} classifier, where the largest improvement of $\SI{17.5+-4.7}{\percent}$ over the baseline setup is reached for $\lambda=2$, meaning a training data set consisting of all real images plus twice as many generated images. Additionally, similar improvements are seen for all other $\lambda$ values that have been tested. 

For the considerably more complex \gls{cnn} classifier, the improvement is not so consistent and already the baseline performance is far better than even the enhanced performance of the \gls{fcn} classifier. However, we do obtain a maximal improvement of $\SI{3.0+-1.8}{\percent}$ for $\lambda = 3$, which also represents the overall highest accuracy for any of the setups investigated here.

Finally, for the most complex classifier architecture, the \gls{vit}, we are not able to show a conclusive improvement of the classifier performance, so that we might expect a dependency of the ability of generated images to add useful information to the training data set on the baseline performance (often connected to the complexity) of the classifier in question. A na\"ive interpretation could be that the better performing architectures are simply more sensitive to even small differences between the real and generated images. Additionally, the robustness of the \gls{vit} might be an issue because it was pre-trained with natural images and only fine-tuned with radio galaxy images due to the limited data sample size.

Further, we considered a three-class classification problem with extended sources only. We found that the overall accuracy is reduced as compact sources are easier to classify. More importantly, the significance of the improvement by including generated images in the training is not enhanced as the variations in the cross-validation tend to increase as well.\\

The best overall accuracy is obtained by using the \gls{cnn} and \gls{wgan} augmented training data, but only by a small margin. Yet, we have shown that \gls{wgan} augmentation works in principle (similar to the goal in \citet{GANplify_2021}, as noted above) and can significantly improve a somewhat simpler algorithm. This can be useful for applications of classification algorithms in resource-constrained environments, i.e. disk-space and inference time restrictions.\\

Our generative model is able to generate large sets of radio galaxy images of different morphologies very quickly. A batch of 100 images can be generated on a  NVIDIA V100 GPU in $\mathtt{\sim}0.1~\text{seconds}$  and in $\mathtt{\sim}4.5~\text{seconds}$ on CPU. Therefore, our \gls{wgan} can play an important role in the simulation and analysis of large radio surveys. Future work involving much larger training sets from the LOFAR telescope will explore this further. Moreover, \gls{wgan}-generated images can be used to validate new interferometric machine-learning algorithms, see e.g. \cite{Schmidt_2022}. To this end, we provide the model and weights with documentation at \url{https://github.com/floriangriese/wGAN-supported-augmentation}.

\section*{Acknowledgements}

This work was supported by UHH, DESY, TUHH and HamburgX grant LFF-HHX-03 to the Center for Data and Computing in Natural Sciences (CDCS) from the Hamburg Ministry of Science, Research, Equalities and Districts. This project benefits greatly from the exchange with particle physicists with a vast experience in using generative models for calorimeter simulations and was supported in part through the Maxwell computational resources operated at DESY. We acknowledge financial support from the Open Access Publication Fund of Universität Hamburg. 

\section*{Data Availability}
\label{sec:DataAvail}
The code of all models trained for this work is publicly available on GitHub at the following address: \url{https://github.com/floriangriese/wGAN-supported-augmentation} \newline
The data set is available on Zenodo at \url{https://doi.org/10.5281/zenodo.7120632} and code for data loading on GitHub at \url{https://github.com/floriangriese/RadioGalaxyDataset}. If you use this data set, please cite \cite{GRIESE2023108974}. \newline


\FloatBarrier



\bibliographystyle{rasti}
\bibliography{lit} 




\appendix
\section{Generated and real images}
\label{App:img}
\begin{figure*}
    \centering
    \includegraphics[width=1.0\textwidth,trim={3cm 8cm 3cm 8cm},clip]{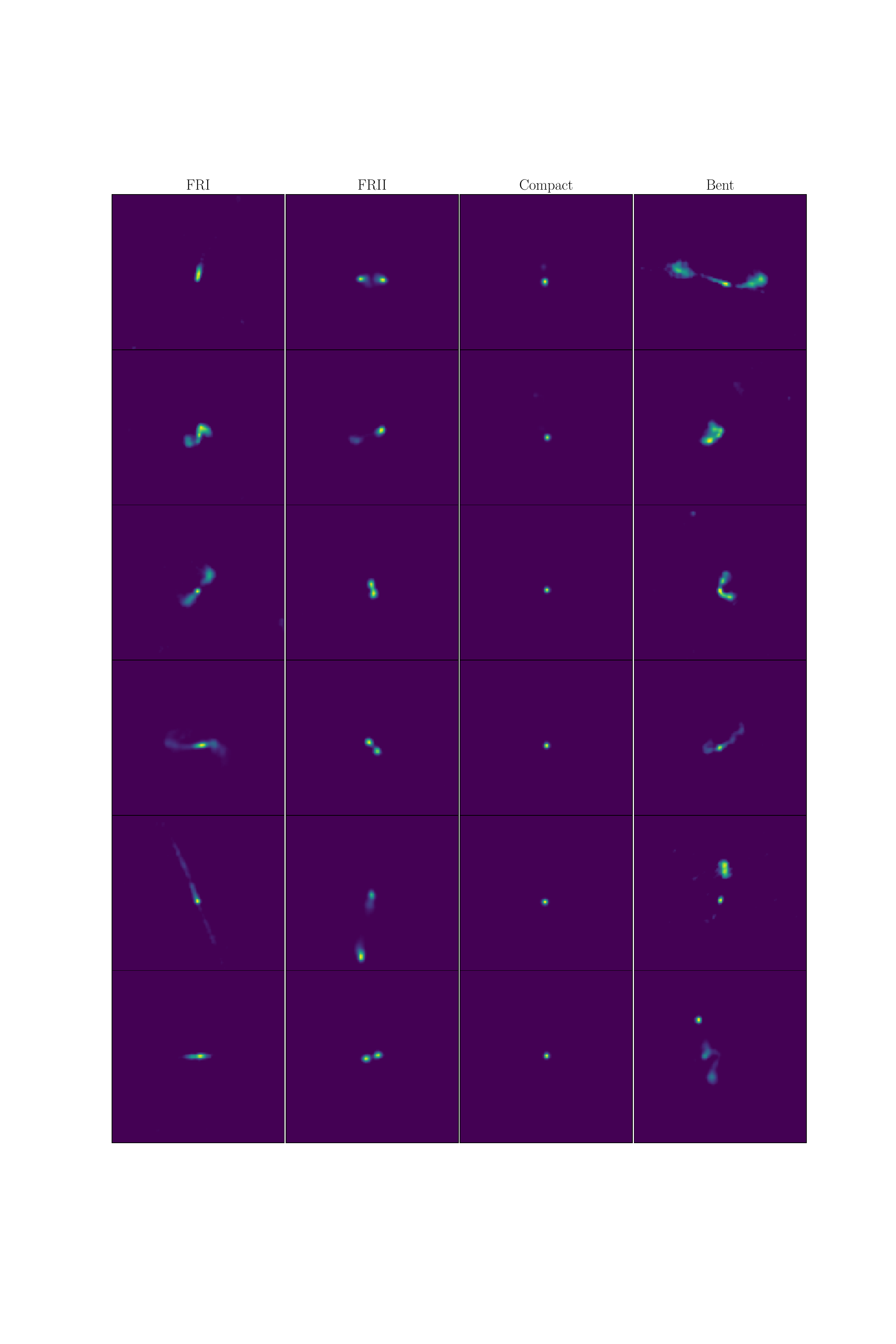}
    \caption{Real examples for each class.}
    \label{fig:real_grid}
\end{figure*}

\begin{figure*}
    \centering
    \includegraphics[width=1.0\textwidth,trim={3cm 8cm 3cm 8cm},clip]{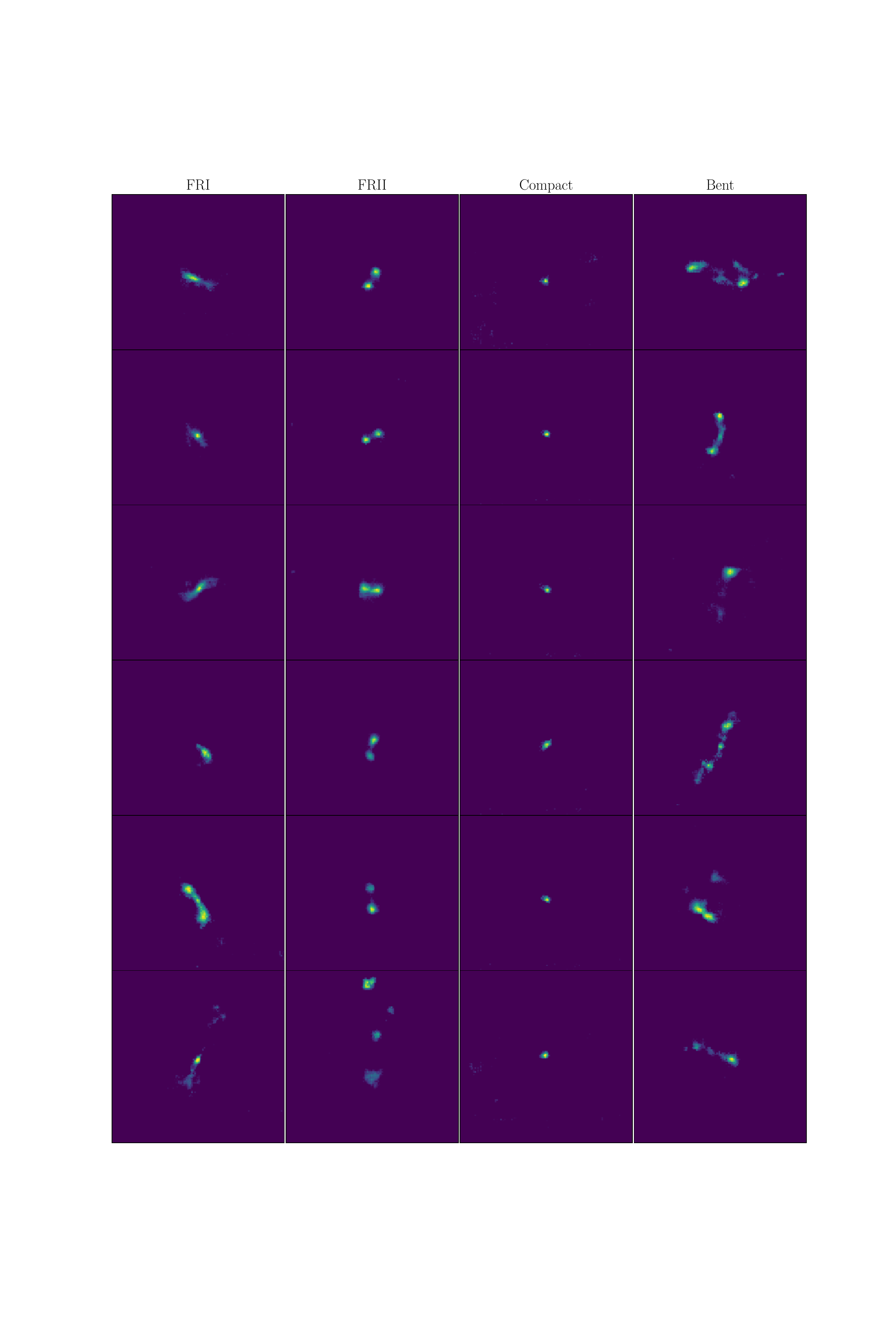}
    \caption{Generated examples for each class.}
    \label{fig:gen_grid}
\end{figure*}

Here we show a random sample of 24 real images in \autoref{fig:real_grid} and 24 generated images in \autoref{fig:gen_grid} in order to give a visual impression of the diversity of the data. 

\section{Classifier architectures}
Detailed information about the architecture of the implemented models is given in this appendix. The structure and the corresponding number of parameters for the the critic of the \gls{wgan} is given in \autoref{tab:wgan_critic}, and for the generator in 
\autoref{tab:wgan_generator}. Detailed information about the \gls{cnn} is given in \autoref{tab:cnn_classifier} and for the \gls{fcn} in \autoref{tab:fcn_classifier}. In \autoref{tab:hyperparam} we summarise the hyperparameters of all model trainings we conducted for this study. 

\begin{table*}
    \centering
    \caption{Parameters of the \gls{wgan} critic.}
    \label{tab:wgan_critic}
    \begin{tabular}{ccccccccc}
         \toprule
         \text{Layer} & \text{Name} & \text{Kernel size} & \text{stride} & \text{Input channels} & \text{Depth} & \text{Activation} &\text{Regularizer} & \text{Parameters}  \\ \midrule
         1 & \text{Conv1} & 4 x 4 & 2 & 1 & 32 & \text{Leaky ReLU} & \text{Layer Norm} & 512 \\
         2 & \text{Conv2} & 4 x 4 & 2 & 4 & 32 & \text{Leaky ReLU} & \text{Layer Norm} & 2,048 \\
         3 & \text{Conv3} & 4 x 4 & 2 & 64 & 128 & \text{Leaky ReLU} & \text{Layer Norm} & 133,120 \\ 
         4 & \text{Conv4} & 4 x 4 & 2 & 64 & 256 & \text{Leaky ReLU} & \text{Layer Norm} & 524,800 \\
         5 & \text{Conv5} & 4 x 4 & 2 & 256 & 512 & \text{Leaky ReLU} & \text{Layer Norm} & 2,097,280 \\
         6 & \text{Conv6} & 4 x 4 & 2 & 512 & 1024 & \text{Leaky ReLU} & \text{Layer Norm} & 8,388,640 \\
         7 & \text{Conv7} & 4 x 4 & 1 & 1024 & 1 & - & - & 16,384 \\
         \midrule
         \text{Total parameters}:& & & & & & & & 11,162,784 \\ \bottomrule
    \end{tabular}
\end{table*}

\begin{table*}
    \centering
    \caption{Parameters of the \gls{wgan} generator.}
    \label{tab:wgan_generator}
    \begin{tabular}{ccccccccc}
         \toprule
         \text{Layer} & \text{Name} & \text{Kernel size} & \text{stride} & \text{Input channels} & \text{Depth} &  \text{Activation} &\text{Regularizer} & \text{Parameters}  \\ \midrule
         1 & \text{ConvT1} & 4 x 4 & 1 & 100 & 512 & \text{ReLU} & \text{Batch Norm} & 820,224 \\
         2 & \text{ConvT2} & 4 x 4 & 1 & 4 & 512 & \text{ReLU} & \text{Batch Norm} & 33,792 \\
         3 & \text{ConvT3} & 4 x 4 & 2 & 1024 & 512 & \text{ReLU} & \text{Batch Norm} & 8,389,632 \\ 
         4 & \text{ConvT4} & 4 x 4& 2 & 512 & 256 & \text{ReLU} & \text{Batch Norm} & 2,097,664 \\
         5 & \text{ConvT5} & 4 x 4 & 2 & 256 & 128 & \text{ReLU} & \text{Batch Norm} & 524,544 \\
         6 & \text{ConvT6} & 4 x 4 & 2 & 128 & 64 & \text{ReLU} & \text{Batch Norm} & 131,200 \\
         7 & \text{ConvT7} & 4 x 4 & 2 & 64 & 1 & - & - & 1,024 \\
         \midrule
         \text{Total parameters}:& & & & & & & & 11,998,080 \\ \bottomrule
    \end{tabular}
\end{table*}

\begin{table*}
    \centering
    \caption{Parameters of the \acrlong{cnn} classifier.}
    \label{tab:cnn_classifier}
    \begin{tabular}{ccccccccc}
         \toprule
         \text{Layer} & \text{Name} & \text{Kernel size} & \text{stride} & \text{Input channels} & \text{Depth} & \text{Activation} &\text{Regularizer} & \text{Parameters}  \\ \midrule
         1 & \text{Conv1} & 3 x 3 & 2 & 1 & 8 & \text{Leaky ReLU} & \text{Layer Norm} & 8264 \\
         2 & \text{Conv2} & 3 x 3 & 2 & 8 & 16 & \text{Leaky ReLU} & \text{Layer Norm} & 3200 \\ 
         3 & \text{Conv3} & 3 x 3 & 2 & 16 & 32 & \text{Leaky ReLU} & \text{Layer Norm} & 5120 \\
         4 & \text{Conv4} & 3 x 3 & 2 & 32 & 32 & \text{Leaky ReLU} & \text{Layer Norm} & 9344 \\
         5 & \text{Conv5} & 2 x 2 & 1 & 32 & 16 & \text{Leaky ReLU} & - & 2048 \\
         6 & \text{Fully Connected 1} & & & 7 x 7 x 16 & 100 & \text{Leaky ReLU} & - & 78500 \\
         7 & \text{Fully Connected 2} & & & 100 & 4 & \text{ReLU} & - & 404 \\
         8 & \text{Softmax} & & & & & & & \\
         \midrule
         \text{Total parameters}:& & & & & & & & 106,880 \\ \bottomrule
    \end{tabular}
\end{table*}

\begin{table*}
    \centering
    \caption{Parameters of the \acrlong{fcn} classifier.}
    \label{tab:fcn_classifier}
    \begin{tabular}{cccccc}
         \toprule
         \text{Layer} & \text{Name} & \text{Input channels} & \text{Depth} & \text{Activation} & \text{Parameters}  \\ \midrule
         1 & \text{Fully Connected 1}  & 128 x 128 & 250 & \text{Leaky ReLU} & 4,096,250 \\
         2 & \text{Fully Connected 2}  & 250 & 250 & \text{Leaky ReLU} & 62,750 \\ 
         3 & \text{Fully Connected 3}  & 250 & 250 & \text{Leaky ReLU} &  62,750 \\
         4 & \text{Fully Connected 4}  & 250 & 250 & \text{Leaky ReLU} &  62,750 \\
         5 & \text{Fully Connected 5}  & 250 & 4 & 
         & 1,004 \\
         6 & \text{Softmax} & & & & \\
         \midrule
         \text{Total parameters}:& & & & & 4,285,504 \\ \bottomrule
    \end{tabular}
    
\end{table*}

\begin{table*}
    \centering
    \caption{Hyperparameters of the trainings.}
    \label{tab:hyperparam}
    \begin{tabular}{ccccccccc}
         \toprule
          & \text{Batch Size} & \text{Learning Rate} & \text{Optimizer} & $\beta_1$ & $\beta_2$ & \text{Momentum} & Training time & Iterations \\ \midrule
         \text{\gls{wgan}} & 400  & 0.0001 & \text{Adam} & 0 & 0.9 & - & $\approx7$h& 40k\\
         \text{Classifier (\gls{fcn} \& \gls{cnn})}& 250 & 0.001 & \text{Adam} & 0.9 & 0.999 & - & 72h & $\approx 300$k\\ 
         \text{\gls{vit}}& 32 & 0.03 & \text{SGD} & - & - & 0.9 & $\approx 8$h & 10k \\ 
         \bottomrule
    \end{tabular}
    
\end{table*}

\newpage

\section{Class-wise performance}
\label{App:AddPlots}
\begin{figure}
\vspace{1cm}
    \centering
    \includegraphics[width=.5\textwidth]{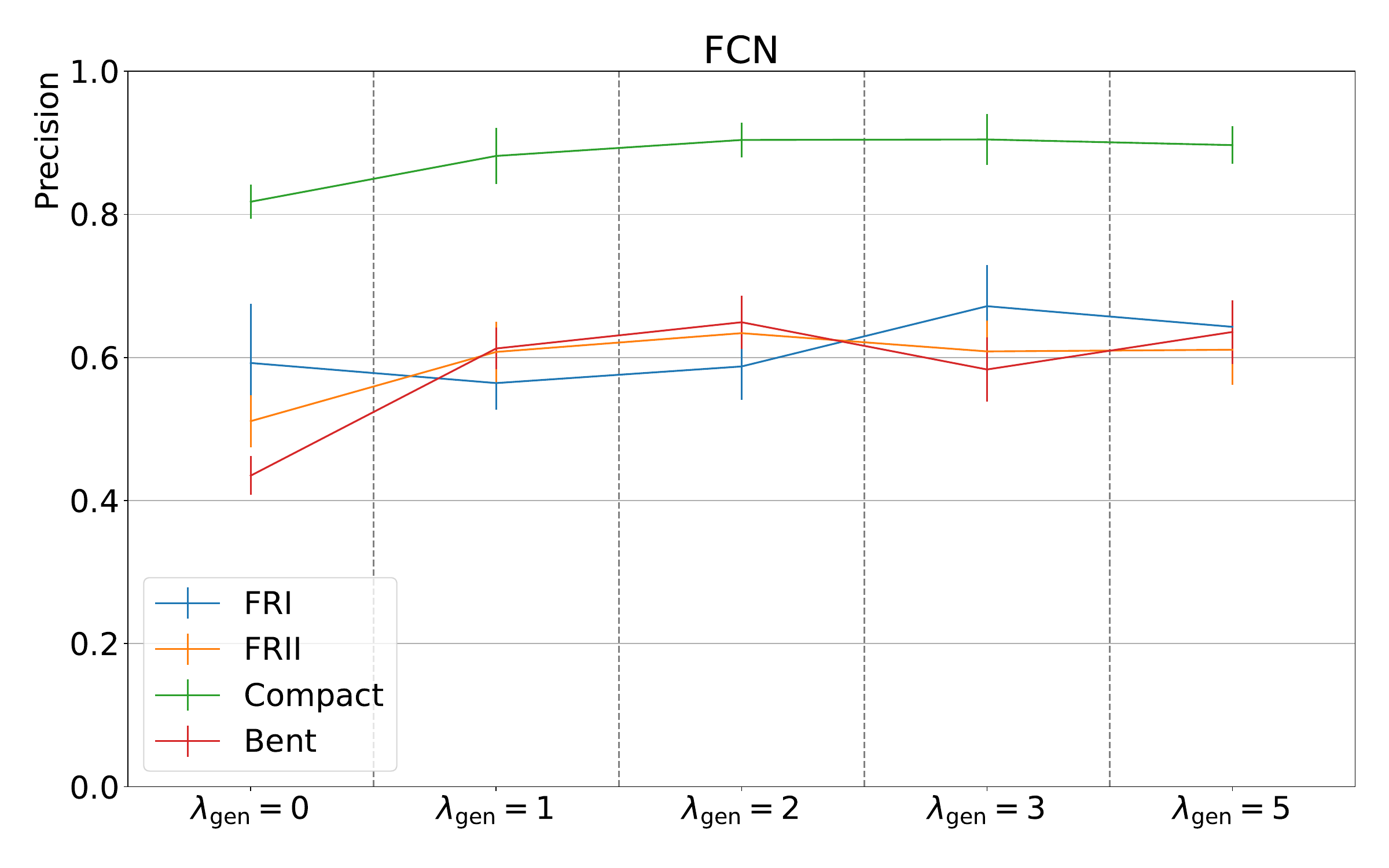}
    \includegraphics[width=.5\textwidth]{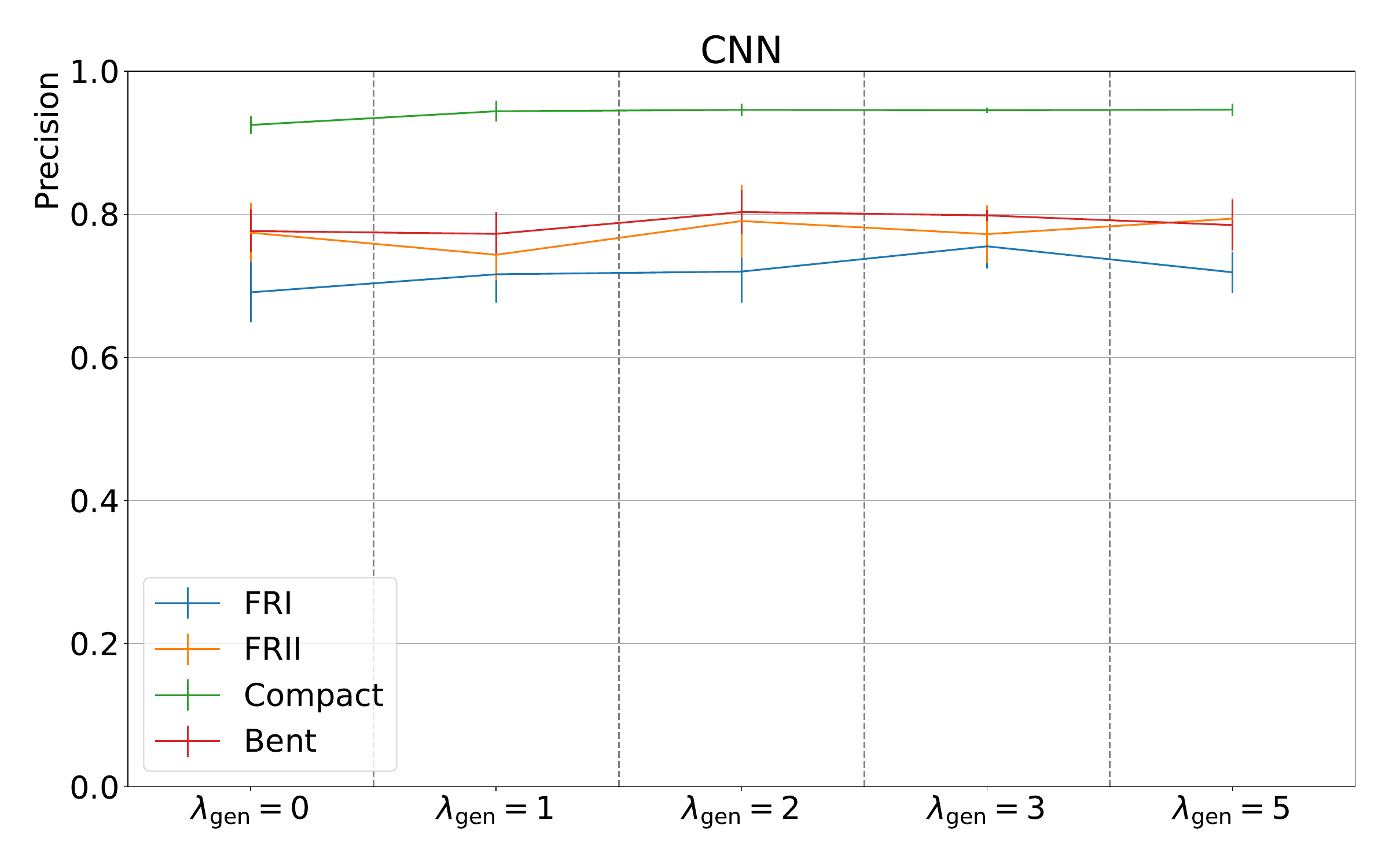}
    \includegraphics[width=.5\textwidth]{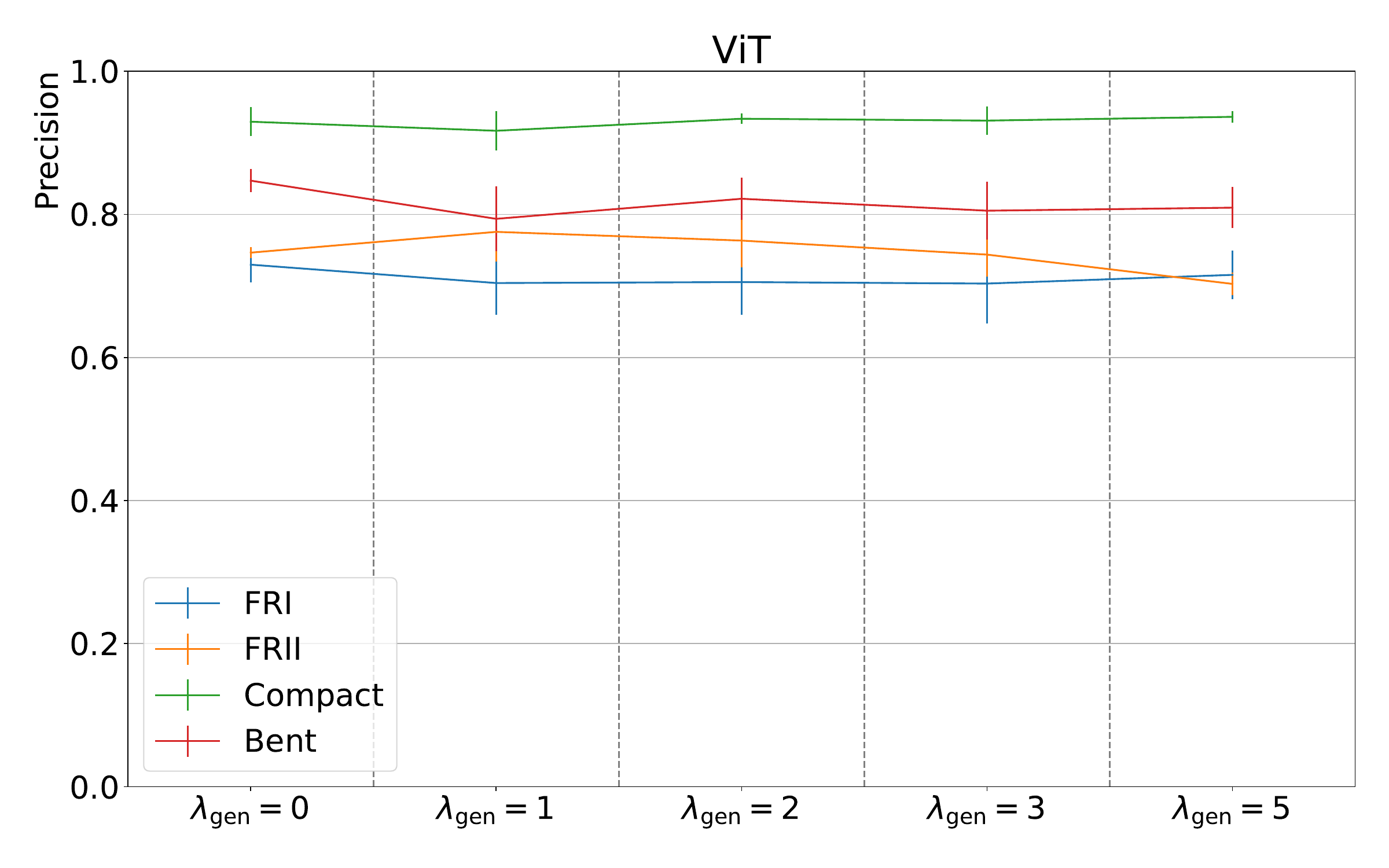}
    
    \caption{Precision for each class on the test data set for the three different classifier architectures for different training scenarios. The markers (uncertainty bars) represent the mean (standard deviation) of the obtained results over all cross-validation folds for the classically + \gls{wgan} augmented training data sets.}%
    \label{Fig:precison}
\end{figure}

\begin{figure}
\vspace{1cm}
    \centering
    \includegraphics[width=.5\textwidth]{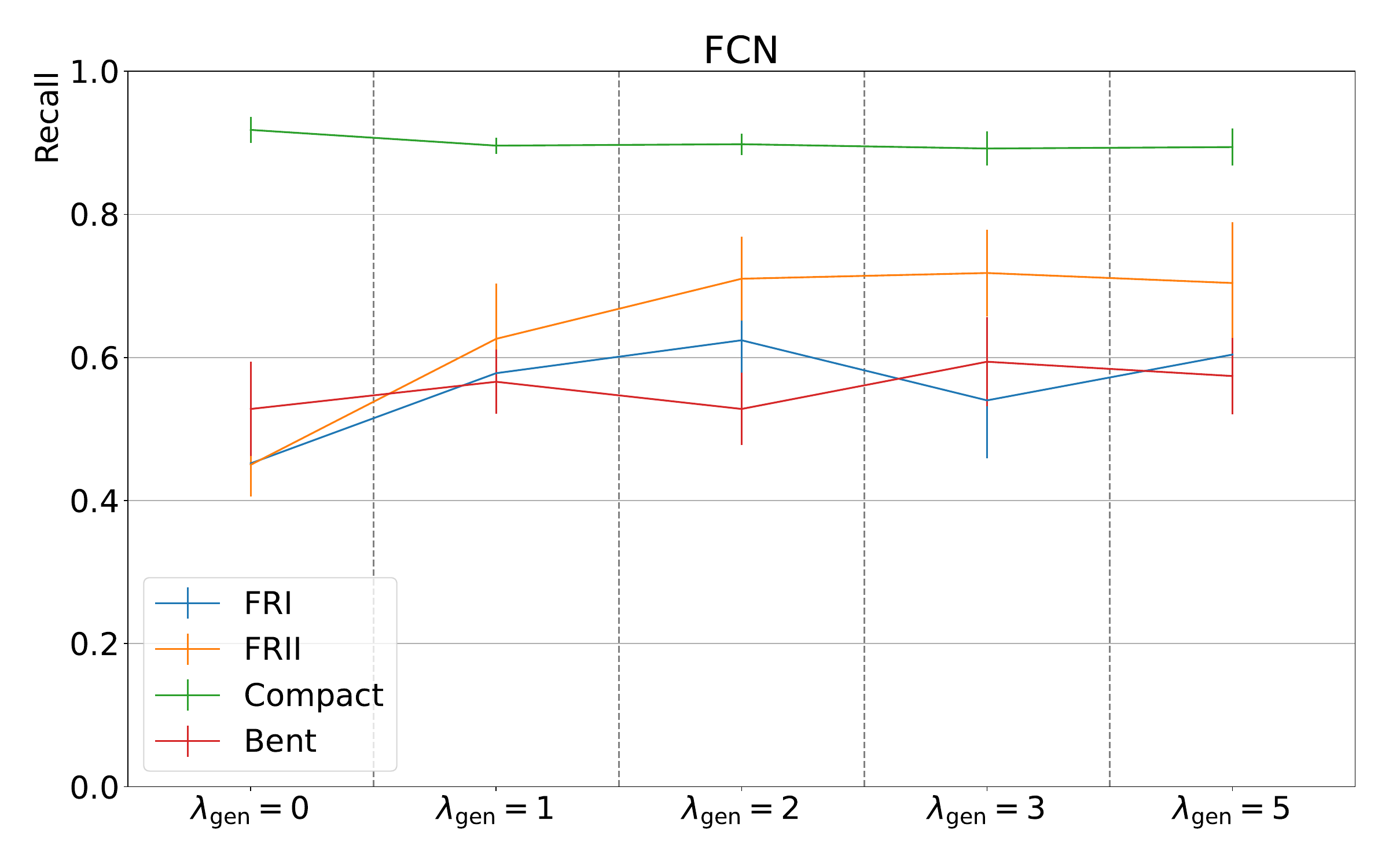}
    \includegraphics[width=.5\textwidth]{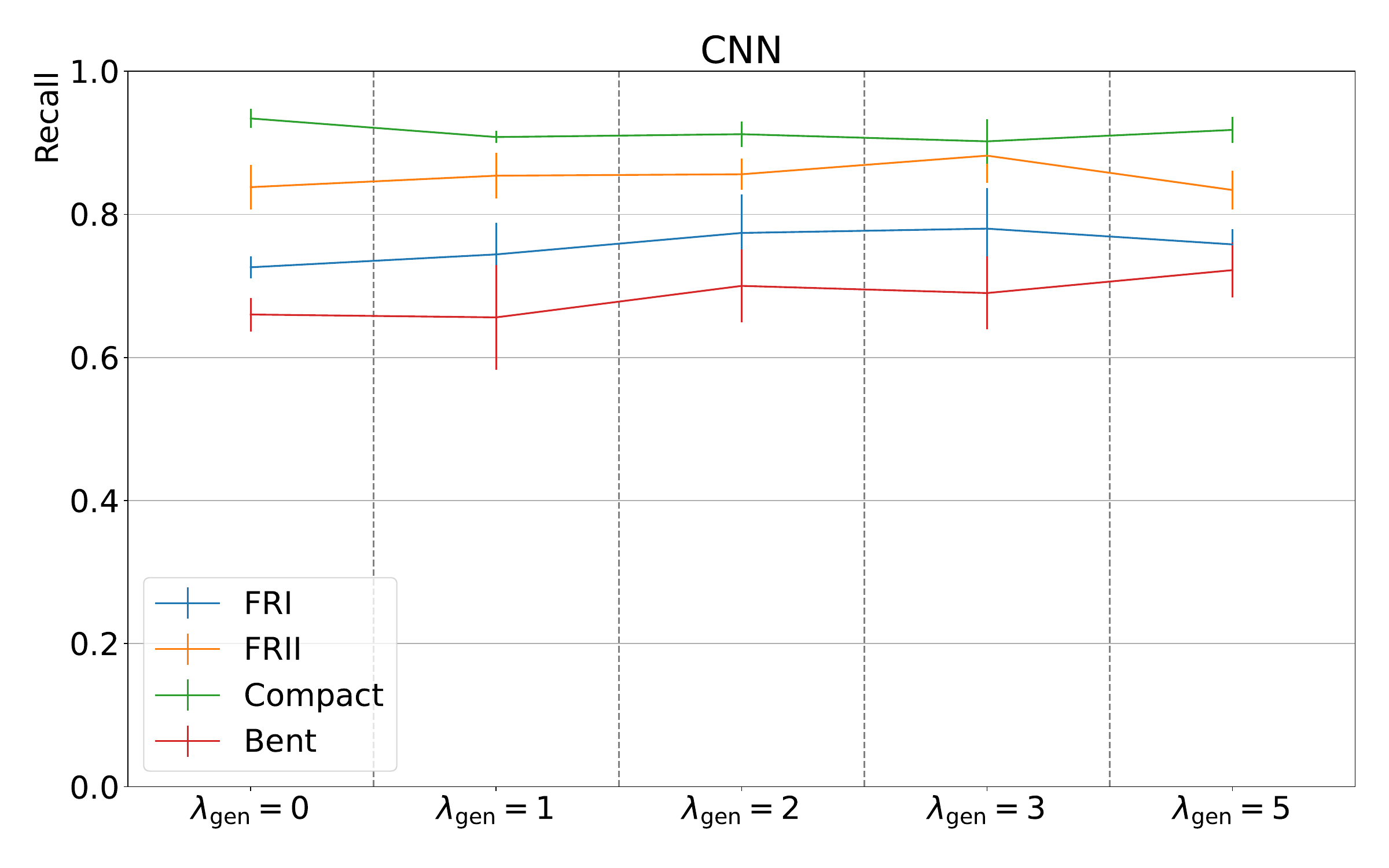}
    \includegraphics[width=.5\textwidth]{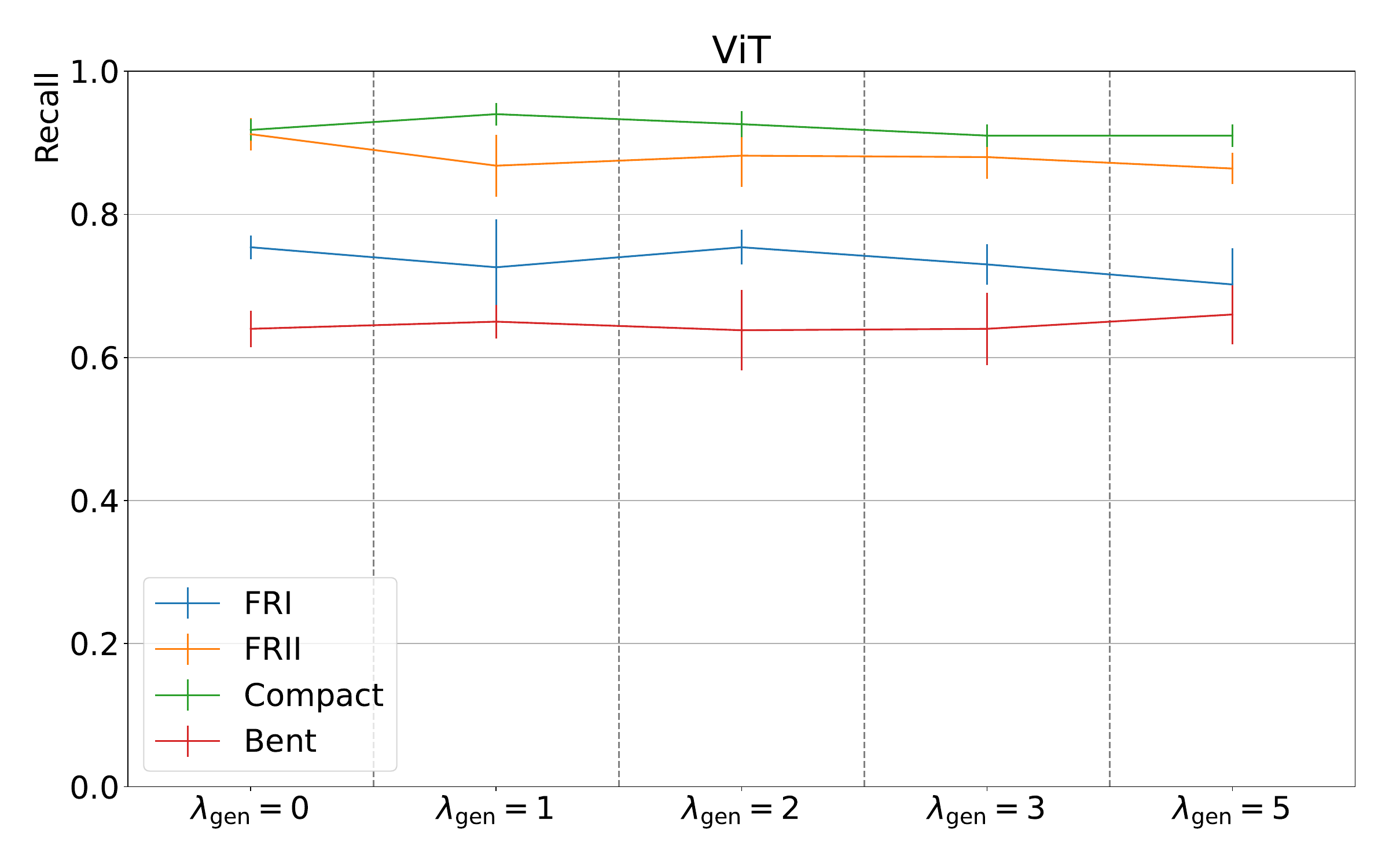}
    \caption{Recall for each class on the test data set for the three different classifier architectures for different training scenarios}%
    \label{Fig:recall}
\end{figure}

\begin{figure}
\vspace{1cm}
    \centering
    \includegraphics[width=.5\textwidth]{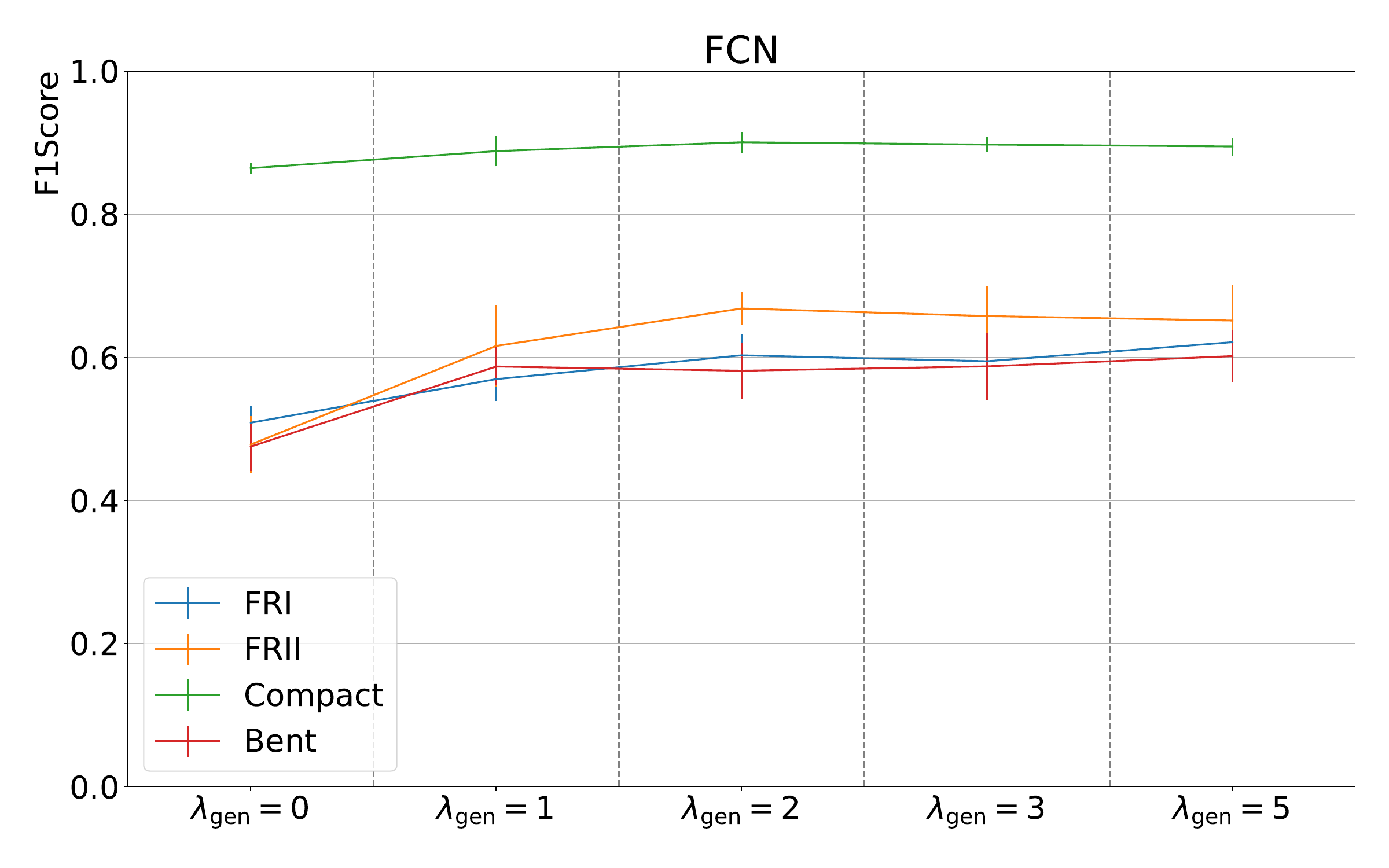}
    \includegraphics[width=.5\textwidth]{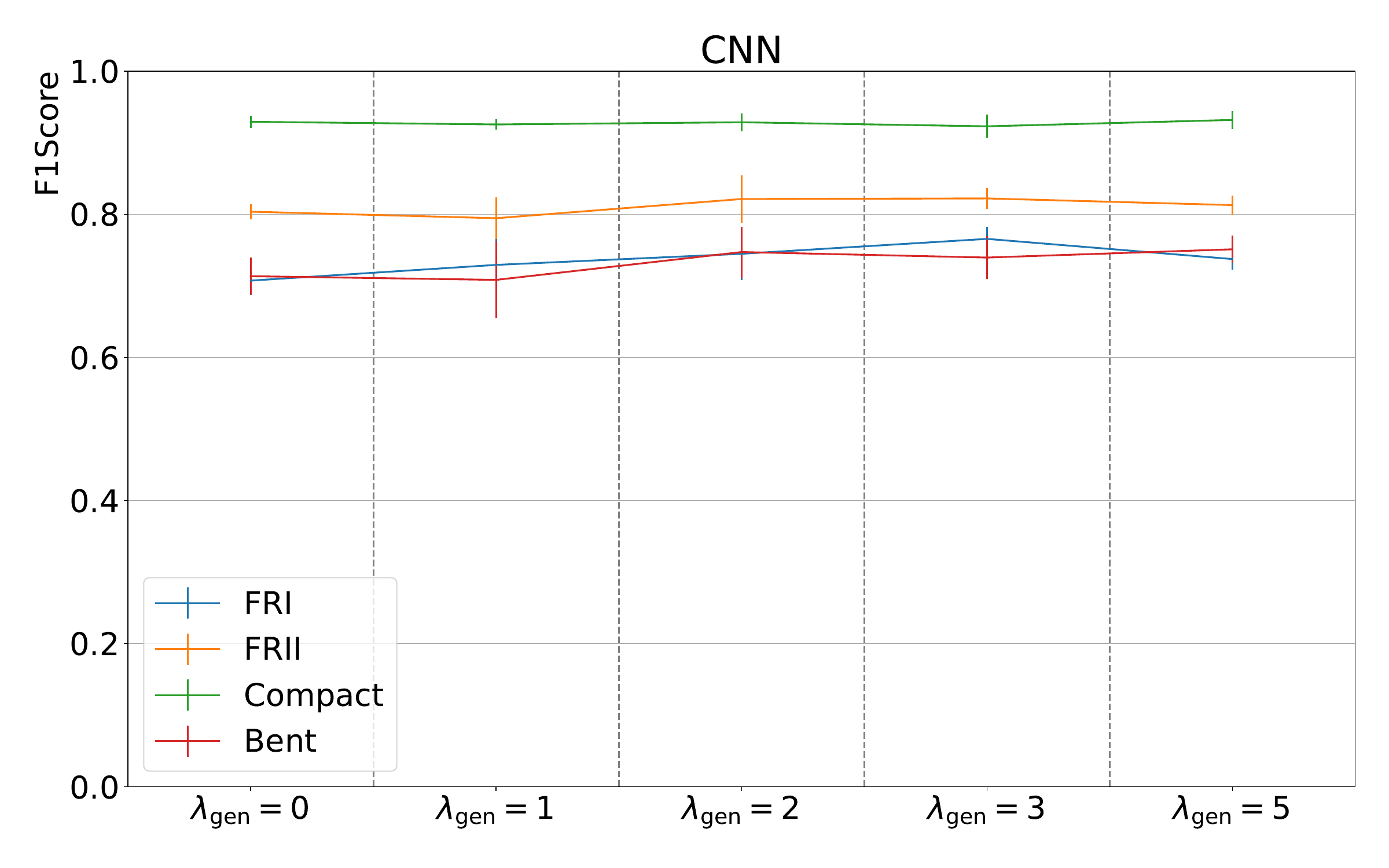}
    \includegraphics[width=.5\textwidth]{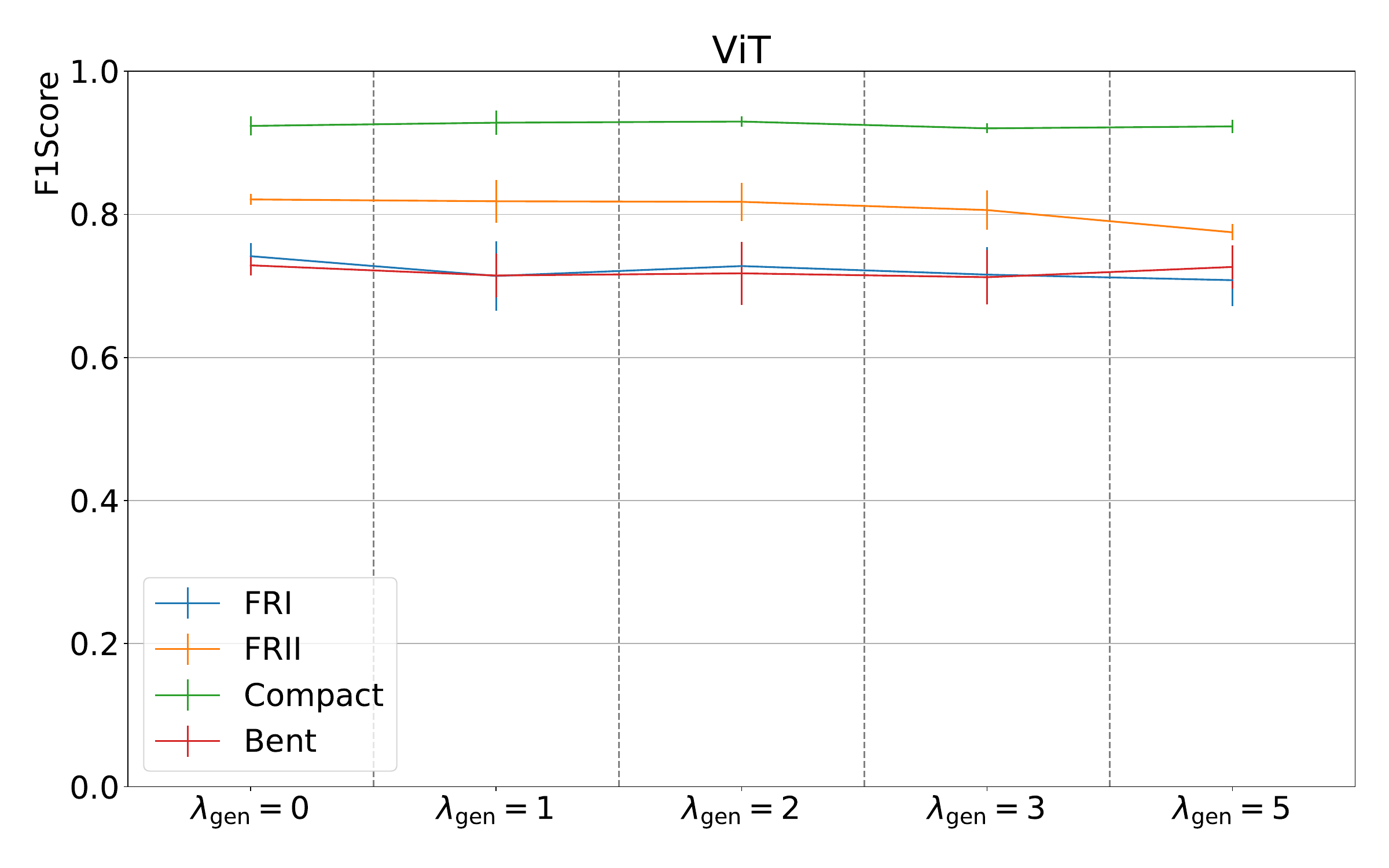}
    \caption{F1 Score for each class on the test data set for the three different classifier architectures for different training scenarios}%
    \label{Fig:f1}
\end{figure}

In this section we demonstrate the performance per class of the classifiers studied in \autoref{sec:wGANsupAug}. In particluar, we show the class-wise Precision in \autoref{Fig:precison}, Recall in \autoref{Fig:recall} and F1 Score in \autoref{Fig:f1}.

\newpage
\ 
\newpage 

\section{Additional Test}
\label{App:AddTest}
\begin{figure}
\vspace{1cm}
    \centering
    \includegraphics[width=.5\textwidth]{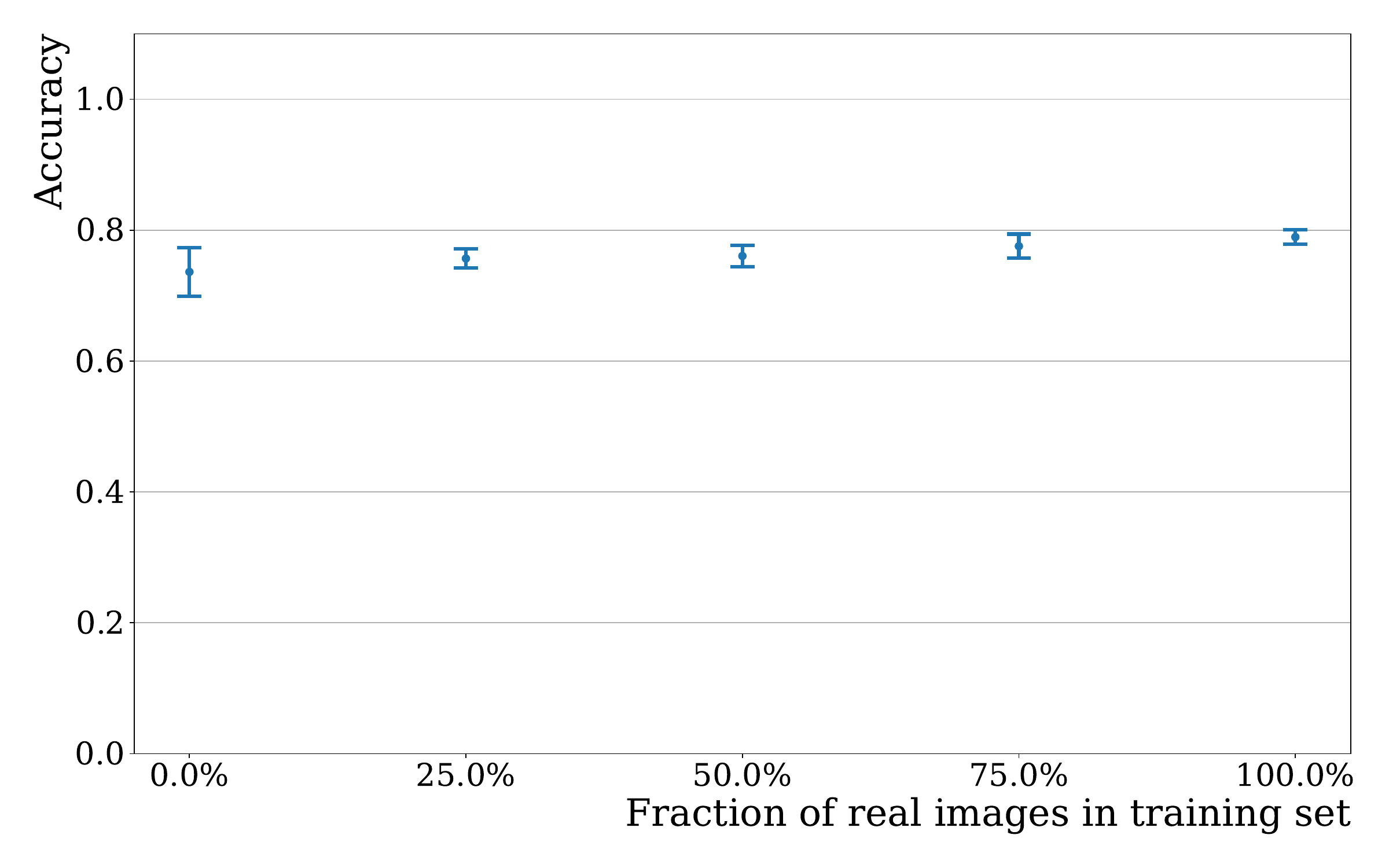}
    \caption{Accuracy on the test set achieved by the best CNN model trained on combined, i.e. generated + real, data sets with varying fractions of real images. The overall number of images in the training set corresponds to the full real-only training set and class imbalance is kept. Classical augmentation is used on both types of images during training.}%
    \label{fig:add_test}
\end{figure}

Here we present an additional test to compare the information content of real and generated images during classifier training. The classifier architecture for this test is the \gls{cnn} introduced in \autoref{tab:cnn_classifier}. We train the \gls{cnn} on different compositions of the original training set and a batch of generated images of the same size. We observe that the classifier performance worsens gradually as we remove real images and add generated images to keep the size of the training set fixed (see \autoref{fig:add_test}). From this experiment we can confidently conclude that the generated images are less informative compared to real images. Note, that we had to exclude some runs with low amount of real data due to the inability to classify the compact sources correctly along with the extended sources. 

\bsp	
\label{lastpage}
\end{document}